\documentclass[twoside,twocolumn]{article}

\usepackage{blindtext} % Package to generate dummy text throughout this template 

\usepackage[sc]{mathpazo} % Use the Palatino font
\usepackage[T1]{fontenc} % Use 8-bit encoding that has 256 glyphs
\usepackage{microtype} % Slightly tweak font spacing for aesthetics
\usepackage[english]{babel} % Language hyphenation and typographical rules

\usepackage[hmarginratio=1:1,top=32mm,columnsep=20pt]{geometry} % Document margins
\usepackage[hang, small,labelfont=bf,up,textfont=it,up]{caption} % Custom captions under/above floats in tables or figures
\usepackage{booktabs} % Horizontal rules in tables
\usepackage{dblfloatfix}    % To enable figures at the bottom of page
\usepackage{lettrine} % The lettrine is the first enlarged letter at the beginning of the text

\usepackage{enumitem} % Customized lists

\usepackage{abstract} % Allows abstract customization
\usepackage{titlesec} % Allows customization of titles
\renewcommand\thesection{\Roman{section}} % Roman numerals for the sections
\renewcommand\thesubsection{\roman{subsection}} % roman numerals for subsections
\titleformat{\section}[block]{\large\scshape\centering}{\thesection.}{1em}{} % Change the look of the section titles
\titleformat{\subsection}[block]{\large}{\thesubsection.}{1em}{} % Change the look of the section titles
\usepackage{afterpage}
\usepackage{fancyhdr} % Headers and footers
\usepackage{titling} % Customizing the title section

\usepackage{hyperref} % For hyperlinks in the PDF
\usepackage{natbib}
\usepackage[pdftex]{graphicx}     

\pretitle{\begin{center}\Huge\bfseries} % Article title formatting
\posttitle{\end{center}} % Article title closing formatting
\title{How Does the Antarctic Circumpolar Current Affect the Southern Ocean Meridional Overturning Circulation?} % Article title
\author{%
\textsc{Christopher C. Chapman}\thanks{\textit{Corresponding author address:} 
				C. C. Chapman, LOCEAN-IPSL, 
				Universit\'{e} de Pierre et Marie Curie, Paris CEDEX ,France. 
				\newline{E-mail: chris.chapman.28@gmail.com}} \\[1ex] % Your name
\normalsize LOCEAN-IPSL\\ Universit\'{e} de Pierre et Marie Curie \\ % Your institution
\normalsize \href{mailto:chris.chapman.28@gmail.com}{mailto:chris.chapman.28@gmail.com} % Your email address
\and % Uncomment if 2 authors are required, duplicate these 4 lines if more
\textsc{Jean-Baptiste Sall\'{e}e}\thanks{Corresponding author} \\[1ex] % Second author's name
\normalsize LOCEAN-IPSL\\ CNRS/Universit\'{e} de Pierre et Marie Curie \\ % Your institution
}
\date{\today} % Leave empty to omit a date
\renewcommand{\maketitlehookd}{%
\begin{abstract}
\noindent The Meridional Overturning Circulation (MOC) in the Southern Ocean is investigated using hydrographic observations combined with satellite observations of sea-surface height. A three-dimensional (spatial and vertical) estimate of the isopycnal eddy-diffusivity in the Southern Ocean is obtained using the theory of Ferrari \& Nikurashin (2010), that includes the influence of suppression of the diffusivity by the strong, time-mean flows. It is found that the eddy diffusivity is enhanced at depth, reaching a maximum at the ``critical layer'' near 1000m. 
The estimate of diffusivity is used with a simple diffusive parameterization to estimate the meridional eddy volume flux. Together with an estimate of the meridional Ekman transport and the time-mean meridional geostrophic transport, the eddy volume flux is used to reconstruct the time-mean overturning circulation. By comparing the reconstruction with, and without, suppression of the eddy diffusivity by the mean flow, the influence of the suppression on the overturning is illuminated. It is shown that the suppression of the eddy diffusivity results in a large reduction of interior eddy transports, and a more realistic eddy induced overturning circulation. Finally, a simple conceptual model is used to show that the MOC is influenced not only by the existence of enhanced diffusivity at depth, but also by the details of the vertical structure of the eddy diffusivity, such as the depth of the critical layer.   % Dummy abstract text - replace \blindtext with your abstract text
\end{abstract}
}

\begin{document}

% Print the title
\maketitle

%----------------------------------------------------------------------------------------
%	ARTICLE CONTENTS
%----------------------------------------------------------------------------------------

\section{Introduction}

\lettrine[nindent=0em,lines=3]{T}he Meridional Overturning Circulation (MOC), is a global scale circulation that, due to its important role in the redistribution of heat, salt and biogeochemical tracers from warmer to colder latitudes, and in the subduction of atmospheric CO$_2$, has a large influence on the climate system \citep{TalleyEtAl2003,Marshall&Speer2012}. In the Southern Ocean, the overturning is related to the rate that  deep carbon-rich waters are ventilated at the surface where they can communicate with the atmosphere, and the rate at which surface waters are in turn subducted into the ocean interior \citep{SalleeEtAl2013}. Thus, changes in the rate of the overturning have been hypothesized to lead to a reduction in the Southern Ocean's ability to absorb and sequester CO$_2$ \citep{LeQuereEtAl2007}. Understanding the dynamic controls of the MOC in the Southern Ocean, as well as how it will respond to external changes in the climate system, is therefore a pressing question in physical oceanography.  

Motivated by the widely acknowledged importance of the Southern Ocean for the global MOC, intense focus on this region has led to significant advances in our understanding of the structure of the MOC and dominant dynamical mechanisms that lead to its formation. In particular, a description of the Southern Ocean based on the \textit{Transformed Eulerian Mean} (TEM) formulation has shown that, on the large scale, the Southern Ocean overturning results from a competition between a northward wind driven Eulerian mean overturning cell, $\overline{\Psi}$, and a southward eddy induced overturning, $\Psi^{\star}$ \citep{Johnson&Bryden1989,Doos&Webb1994}. The mean and eddy-induced overturning are thought to be of similar magnitude, yet opposite sign, such that only a small residual transport remains. The resulting overturning, commonly expressed as an overturning streamfunction, is written:
\begin{equation} \label{residual_overturning}
\Psi^{\textrm{res}} = \overline{\Psi} + \Psi^{\star}.
\end{equation}     
Because of the delicate balance between the eddy and mean overturning, the  residual overturning is sensitive to even small changes in one or the other component resulting from changes in surface forcing \citep{Viebahn&Eden2010,AbernatheyEtAl2011,MeredithEtAl2012,Downes&Hogg2013}.   

The Southern Ocean also hosts the Antarctic Circumpolar Current (ACC), a system of currents that are among strongest on Earth. Although the ACC is primarily zonally oriented, it has direct and indirect roles in shaping the Southern Ocean MOC. For instance, the interaction of the ACC with bathymetry results in a significant, but frequently ignored, geostrophic interior overturning circulation that is distinct from the mean ageostrophic overturning associated with Ekman currents \citep{MacCready&Rhines2001,Mazloff2008,Mazloff2013}. In addition, the ACC has a strong influence on the stirring ability of meso-scale eddies and hence their ability of move water poleward \citep{BatesEtAl2014}. The capacity of eddies to induce a downgradient flux is often measured by the eddy diffusivity, $K$, which relates the eddy-flux of some tracer with concentration $C$, to the large scale gradient of that tracer:
\begin{equation} \label{diffusivity_def}
\overline{\mathbf{u}^{\prime}C^{\prime}} = K \nabla C
\end{equation}     
Although baroclinic eddies are ubiquitous in the Southern Ocean, certain regions, referred to as ``hot-spots" or ``storm tracks", which arise from the interaction of the ACC with bathymetry \citep{WilliamsEtAl2007,ChapmanEtAl2015}, cause localised increases in $K$ \citep{SalleeEtAl2008}. The ACC also modulates the vertical structure of eddy diffusivity, which is known to be enhanced at depth, reaching a maxima at the ``steering-level" or ``critical layer depth" \citep{Ferrari&Nikurashin2010,Klocker&Abernathey2014}, associated with the fastest growing linear waves \citep{Smith&Marshall2009}. Linear stability analysis places this level at around 1000~m depth \citep{Smith&Marshall2009}. Although it has been shown that including a spatially varying $K$ can reduce bias in coarse resolution climate models \citep{FerreiraEtAl2005,Danabasoglu&Marshall2007}, and while the implications of a three-dimensional $K$ on the broad scale flow have been briefly discussed in several studies \citep{MarshallEtAl2006,ShuckburghEtAl2009a,Smith&Marshall2009,NaveiraGarabatoEtAl2011,BatesEtAl2014}, a detailed understanding of the physical implications of spatially varying $K$ for the large-scale overturning circulation is still lacking.

In this study, we seek to characterize and quantify the influence of ``storm-tracks" and the suppression of the eddy diffusivity by the mean-flow on the Southern Ocean overturning circulation. We will explore the impact of geostrophic mean-flow on the MOC, and in particular the impact of the strong currents of the ACC in modulating the eddy overturning. To achieve our goals, we reconstruct the overturning circulation from a large observational dataset that combines hydrographic data from Argo floats, oceanographic cruises and instrumented elephant seals, with sea surface height altimetry. The observational datasets are used to develop a direct estimate of the Eulerian mean overturing that includes both ageostrophic Ekman currents and the important deep geostrophic currents that arise from the interaction of ACC with the bottom bathymetry. In addition, we produce a three-dimensional estimate of $K$ based on the theory of \cite{Ferrari&Nikurashin2010} that allows a reconstruction of an eddy overturning streamfunction from the downgradient diffusion of potential vorticity \citep{TreguierEtAl1997}. We will investigate the influence of spatial variation of $K$ and the suppressing influence of the background flow on the reconstructed eddy and residual streamfunctions. By reconstructing $\overline{\Psi}$ and $\Psi^{\star}$ from observations, we show that the ACC can impact significantly both of these terms, and has therefore a key role in shaping the residual overturning circulation. In tandem with this reconstruction, we employ a simple conceptual model, based on the TEM approach of \cite{Marshall&Radko2003} and \cite{Marshall&Radko2006} to guide our interpretation of the observational-based reconstruction. 

The remainder of this paper is organized as follows: the theoretical framework used for building the reconstruction of the MOC from observations, including the procedure for estimating the horizontal (isopycnal) diffusivity $K$, will be presented in Section \ref{Section:MOC_Theory}. The observational data set we employ will be described in Section \ref{Section:Data_Methods}. Our estimate of the three-dimensional eddy diffusivity will be presented in Section \ref{Section:K_Calculation} and our estimated MOC reconstruction, along with a comparison of the the results obtained with and without the influence of the background flow in Section \ref{Section:Reconstruction_Results}. The influence of a vertically varying $K$ on the overturning will then be elucidated using a conceptual model in Section \ref{Section:Numerical_Model}. Finally, we will bring together the observational and theoretical results of this study in Section \ref{Sec:Conclusion}.

%------------------------------------------------

\section{The Southern Ocean Meridional Overturning Circulation} \label{Section:MOC_Theory}

Here we briefly revise the basic theory of the MOC in the Southern Ocean, the theory of mean-flow suppression of eddy diffusivity, and the formulation of the TEM model equations. 

On an isopycnal layer, $\gamma$, with thickness $h=-\partial z/\partial \gamma$, the time-mean meridional volume flux is given by:
\begin{equation} \label{Eqn:Meridional_Transport_ReyDecomp}
\overline{hv} = \overline{h}\overline{v} + \overline{h^{\prime}v^{\prime},} 
\end{equation}  
where $v$ is the meridional velocity, $\overline{\left ( \cdot \right )}$ is the time-averaging operator, and the flow has been decomposed into time-mean and eddy components. The primed quantities are perturbations from the time-mean, such that $v=\overline{v}+v^{\prime}$ and $\overline{v^{\prime}}=0$. We can further decompose $v$ into geostrophic, $v_{\textrm{g}}$, and ageostrophic, $v_{\textrm{ag}}$ components:
\begin{equation} \label{Eqn:Meridional_Transport_GeoDecomp}
\overline{hv} = \overline{h}\overline{v_{\textrm{g}}} + \overline{h}\overline{v_{\textrm{ag}}} + \overline{h^{\prime}v_{{\textrm{g}}}^{\prime}} + \overline{h^{\prime}v_{{\textrm{ag}}}^{\prime}}.  
\end{equation}  
The meridional transport can then be vertically integrated on across isopycnal layers to determine the  time-mean isopycnal overturning streamfunction \citep{Doos&Webb1994}:
\begin{equation} \label{Eqn:Streamfunction_Construction}
\overline{\Psi}^{\textrm{res}}(x,y,\gamma) = \overline{\int_{0}^{\gamma} hv\; d\gamma\;^{\prime}} =   \underbrace{\overline{\Psi}_{\textrm{ag}} + \overline{\Psi}_{\textrm{g}}}_{\textrm{mean}} +  \underbrace{\Psi_{\textrm{g}}^{\star} + \Psi_{\textrm{ag}}^{\star}}_{\textrm{eddy}}.  
\end{equation}  
\subsection{The ageostrophic transport}
In the Southern Ocean, the overwhelming majority of the ageostrophic transport, $\overline{h}\overline{v}_{\textrm{ag}}$ occurs due to surface Ekman currents. The ageostrophic eddy transport, $\overline{h^{\prime}v^{\prime}_{\textrm{ag}}}$, although not completely negligible, is much smaller than the time-mean Ekman transport \citep{Mazloff2013}. Since we are unable to estimate this term from the data used in this study we will not discuss it further, although, for reference, \cite{Mazloff2013} finds a southward transport of approximately 5~Sv contained almost entirely in the surface layers. The time-mean Ekman velocity can be determined from the surface wind stress using the equations for an Ekman spiral \citep[pg. 449]{Dutton1986} with a constant Ekman layer depth, here taken to be 100~m, consistent with observations in the Southern Ocean \citep{Lenn&Chereskin2009}.
%\begin{eqnarray} \label{Eqn:Ekman_0}
%u_{ag} & = & \frac{\sqrt{2}}{\rho_0 f h_{\textrm{Ekman}}}e^{z/h_{\textrm{Ekman}}} \left [\tau^{x}\cos \left(\frac{z}{h_{\textrm{Ekman}}} -\frac{\pi}{4} %\right) - \tau^{y}\sin \left(\frac{z}{h_{\textrm{Ekman}}} -\frac{\pi}{4}\right)  \right], \\ \label{Eqn:Ekman_1}
%v_{ag} & = & \frac{\sqrt{2}}{\rho_0 f h_{\textrm{Ekman}}}e^{z/h_{\textrm{Ekman}}} \left [\tau^{x}\sin \left(\frac{z}{h_{\textrm{Ekman}}} -\frac{\pi}{4} %\right) - \tau^{y}\cos \left(\frac{z}{h_{\textrm{Ekman}}} -\frac{\pi}{4}\right)  \right], \label{Eqn:Ekman_2}
%\end{eqnarray}
%where $\tau^{x}$ and $\tau^{y}$ are the zonal and meridional components of the surface wind stress, and $h_{\textrm{Ekman}}$ is a constant Ekman layer depth, here taken to be 100~m, which is consistent with observations in the Southern Ocean \citep{Lenn&Chereskin2009}.

\subsection{Time-mean geostrophic transport}
The time-mean geostrophic transport, $\overline{h}\overline{v}_{\textrm{g}}$ arises due to the outcropping of the isopycnal surfaces either with the ocean surface or the ocean floor \citep{Ward&Hogg2011,Mazloff2013}. On pressure surfaces, the geostrophic velocity can be determined from hydrography by computing the dynamic height anomaly, which gives an exact geostrophic streamfunction. On an isopycnal layer, there is no exact geostrophic streamfunction \citep{McDougall1989}. However, \cite{McDougall&Klocker2010} have formulated an approximate streamfunction on a neutral density surface \citep{Jackett&McDougall1997},  that can be computed from hydrography. The formulation of this streamfunction takes into account the non-linearity in the equation of state and allows for temperature to vary quadratically with pressure along the neutral surface. The geostrophic velocities are then related to the streamfunction, $M$, by: 
\begin{equation} \label{Eqn:Geostophic_Streamfunction}
f \mathbf{u}_{\textrm{g}} = \mathbf{k} \times \nabla M,
\end{equation}    
where $f=2\Omega_{E}\sin \phi$ is the Coriolis parameter.
%In this study, we will use the \cite{McDougall&Klocker2010} streamfunction to define geostrophic velocities. However, for the ease of illustration, we continue with the Montgomery potential for now.}, so that:: 
%\begin{linenomath*}
%\begin{equation} \label{Eqn:Geostophic_Streamfunction}
%f v_{\textrm{g}} = \frac{\partial M}{\partial x} = \frac{\partial}{\partial x} \left (p + %g\rho_0z \right)  
%\end{equation}    
%\end{linenomath*} 
%where $p=p(x,y,z;t)$ is the pressure on the isopycnal surface and $z=z(x,y;t)$ is its %vertical position. 

The meridional geostrophic transport, $hv_{\textrm{g}}$, does not necessarily integrate to zero around a circumpolar circuit due to outcropping of the isopycnals. In fact, the geostrophic transport is of first order importance to the zonally averaged interior mass balance \citep{MacCready&Rhines2001,Koh&Plumb2004,Mazloff2013}; a point we will return to in Section \ref{Section:Reconstruction_Results}.

\subsection{Geostrophic eddy transport}
The geostrophic eddy transport $\overline{h^{\prime}v_{\textrm{g}}^{\prime}}$ is also found to be of first order importance in a number of studies \citep{AbernatheyEtAl2011,Mazloff2013,DufourEtAl2015}. However, we are unable to estimate this term directly from observations, and hence it must be parameterized. Following \cite{MarshallEtAl1999}, we start by noting that the geostrophic eddy flux of Ertel potential vorticity (PV), $q=\frac{f+\zeta}{h}$, can be written as:
\begin{equation} \label{Eqn:PV_Flux}
\overline{v_{\textrm{g}}^{\prime}q^{\prime}} = -\frac{f}{\overline{h}^2}\overline{h^{\prime}v_{\textrm{g}}^{\prime}},
\end{equation}
assuming planetary geostrophic scaling such that $q \approx f/h$ and $h^{\prime}/\overline{h}<<1$. Thus, the geostrophic eddy volume flux can be written as:
\begin{equation} \label{Eqn:Geo_Eddy_Vol_flux_1}
\overline{h^{\prime}v_{\textrm{g}}^{\prime}} = -\frac{\overline{h}}{\overline{q}}\overline{v^{\prime}_{\textrm{g}}q^{\prime}}.
\end{equation} 
We now employ the simple down-gradient diffusive closure for $\overline{v^{\prime}q^{\prime}}\approx K\partial \overline{q}/\partial y$, described in detail by \cite{TreguierEtAl1997} and discussed in numerous papers thereafter \citep{Killworth1997,MarshallEtAl1999,Wardle&Marshall2000,Malcolm&Marshall2000,Wilson&Williams2004,Plumb&Ferrari2005} to give:
\begin{equation} \label{Eqn:Geo_Eddy_Vol_flux_2}
\overline{h^{\prime}v_{\textrm{g}}^{\prime}} = -K \frac{\overline{h}}{\overline{q}} \frac{\partial \overline{q}}{
\partial y}.
\end{equation}
Eqn. \ref{Eqn:Geo_Eddy_Vol_flux_2} is a crude parameterization of the true eddy fluxes with numerous failings \citep{Malcolm&Marshall2000,Wilson&Williams2004}. However, it has been shown to work well when one is interested only in the large-scale flow \citep{MarshallEtAl1999,Plumb&Ferrari2005,KuoEtAl2005}.

\subsection{Eddy diffusivity $K$}
With the closure for the eddy volume flux in terms of the large scale PV gradient, we are able to reconstruct the geostrophic eddy volume flux with knowledge of the eddy diffusivity $K$, which is known to be dependent on the eddy kinetic energy, the spatial and temporal scales of the meso-scale eddies, and on the state of the background mean flow \citep{Smith&Marshall2009,AbernatheyEtAl2010,Ferrari&Nikurashin2010,Klocker&Abernathey2014,BatesEtAl2014}. \cite{Ferrari&Nikurashin2010}, using the assumptions of simple isotropic turbulence modeled by a white-noise process, derived the following expression for $K$, that takes into the account the effects of the mean-flow:
\begin{equation} \label{Eqn:Eddy_Diffusivity}
\tilde{K} = \frac{K_0}{1+k_{\textrm{eddy}}^{2}\tau_{\textrm{eddy}}^{2} \left [ c_p-\overline{u}(x,y,z) \right ]^{2}},
\end{equation}
where the term $K_0$ represents the eddy diffusivity which is unmodified by the mean-flow, $\tilde{K}$ is the modified eddy diffusivity (sometimes called the \textit{effective diffusivity} or the \textit{suppressed diffusivity} for reasons that will soon become apparent), $k_{\textrm{eddy}}$ is the zonal eddy wavenumber, $\tau_{\textrm{eddy}}$ is eddy decorrelation time scale, and $c_p$ is the eddy phase speed. 

As $\overline{u}\rightarrow c_p$, the denominator of Eqn. \ref{Eqn:Eddy_Diffusivity} approaches unity, which means that $\tilde{K} \rightarrow K_0$. In the case where $\overline{u} \neq c_p$, the denominator of Eqn. \ref{Eqn:Eddy_Diffusivity} is greater than unity and $\tilde{K}<K_0$. For this reason, the denominator of Eqn. \ref{Eqn:Eddy_Diffusivity} is called the ``suppression factor" \citep{Klocker&Abernathey2014}. This reaches its minima at the critical level, i.e. where $\overline{u} = c_p$, which is thought to lie at about 1000~m of depth \citep{Smith&Marshall2009}. The suppression factor also varies throughout the Southern Ocean: tracer diffusivity is suppressed \textit{more} in regions where mean currents are \textit{strong} and \textit{less} where they are  \textit{weak} \citep{Klocker&Abernathey2014}. 

With Eqns. \ref{Eqn:Geo_Eddy_Vol_flux_2} and \ref{Eqn:Eddy_Diffusivity} we can reconstruct the eddy component of the MOC. Investigating the role that the spatial variability and suppression of $K$ play in the MOC is the primary goal of this paper.  

\subsection{Simple conceptual model of the Southern Ocean overturning}
In order to guide our analysis based on the observation-based reconstructed overturning, we will use a simple conceptual model, with the aim to further understand the influence that vertically-variable diffusivity has on the stratification and the overturning (Section \ref{Section:Reconstruction_Results}).

Our model has essentially the same form as that of \cite{Marshall&Radko2003} and \cite{Marshall&Radko2006}. Specifically, we solve the zonally averaged TEM equations in the ocean interior: 
\begin{eqnarray} \label{Eqn:TEM_0}
J_{yz} \left ( \Psi^{\textrm{res}},\overline{b} \right) & = & 0, \\ \label{Eqn:TEM_1}
\Psi^{\textrm{res}}(y,\overline{b}) & = & \overline{\Psi}(y) +  \Psi^{\star}(y,\overline{b}), \\ \label{Eqn:TEM_2}
\Psi^{\star} & = & \left. -K\frac{\partial \overline{b}}{\partial y} \middle / \frac{\partial \overline{b}}{\partial z} \right.,
\end{eqnarray}   
where $b=b(y,z)$ is the buoyancy, $J_{yz}$ is the Jacobian operator, and $\frac{\partial b}{\partial y}  / \frac{\partial b}{\partial z}$ is the isopycnal slope. Here, the mean overturning is taken to be simply that associated with the Ekman flow $\overline{\Psi}=-\tau^{\textrm{wind}}(y)/f$. Rearranging Eqns \ref{Eqn:TEM_1} and \ref{Eqn:TEM_2} and substituting for $\overline{\Psi}$ gives: 
\begin{equation} \label{Eqn:TEM_3}
\left [\frac{\tau^{\textrm{wind}}(y)}{f} + \Psi^{\textrm{res}} \right]\frac{\partial b}{\partial z}  + K\frac{\partial b}{\partial y} = 0.
\end{equation}
Eqn. \ref{Eqn:TEM_3} can be solved numerically using the method of characteristics, as detailed in Appendix B. The boundary conditions are identical to those of \cite{Marshall&Radko2006}: the buoyancy field is set at the base of the homogeneous mixed layer (here taken as $z=0$) and on the northern boundary:
\begin{eqnarray} \label{Eqn:Bdy_Condition_Surface}
b(y,0) = \Delta b^{\textrm{surf}} \frac{y}{L_y},  \\ \label{Eqn:Bdy_Condition_North}
b(L_y,z) = b(L_y,0) e^{\frac{z}{h_e}},
\end{eqnarray}
where $L_y=2000$~km is the meridional scale of the ACC, $h_e=1000~m$ is the e-folding depth, and $\Delta b^{\textrm{surf}}=0.007$~m.s$^{-2}$ is the buoyancy gain across the ACC. Additionally, the surface wind stress, $\tau^{\textrm{wind}}(y)$ is taken to be a simple sinusoidal profile:
\begin{equation} \label{Eqn:wind_stress_def}
\tau^{\textrm{wind}}(y) = \tau_0\left [0.3 +  \sin \left (\frac{\pi}{L_y}y \right ) \right], 
\end{equation}
with $\tau_0=$1.5$\times 10^{-4}$. The residual overturning streamfunction $\Psi^{\textrm{res}}$, is not specified as a surface boundary condition, but is instead calculated as part of the solution using the iterative technique of \cite{Marshall&Radko2006}.

%------------------------------------------------

\section{Observational Data} \label{Section:Data_Methods}
\subsection{Hydrographic Data}
\begin{figure}[t]
   \centering  
  \includegraphics[width=18pc,height=18pc,angle=0]{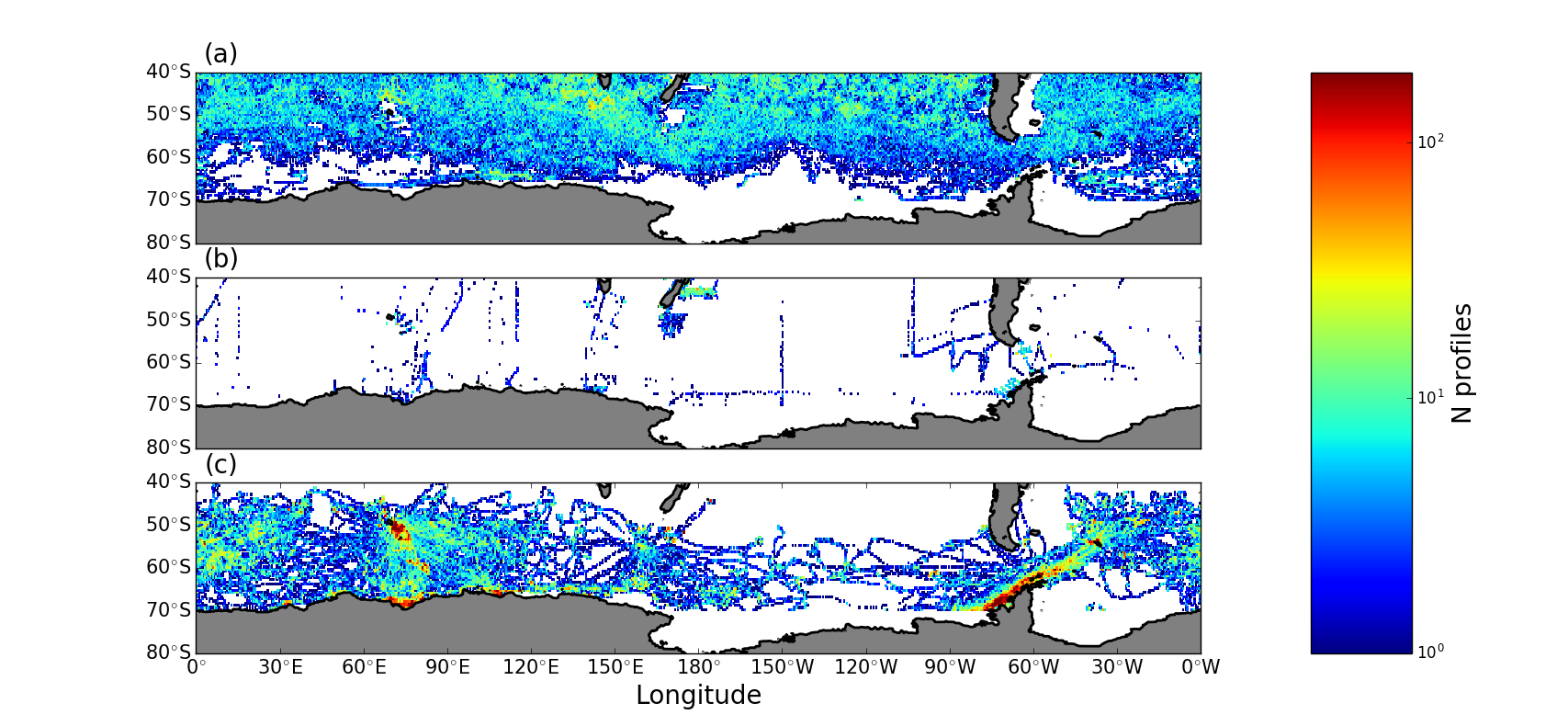}\\
  \caption{Spatial distribution of hydrographic profiles in the Southern Ocean used in this study. The number of (a) Argo profiles ; (b) ship based WOD profiles; and (c) instrumented elephant seals profiles, in each 0.5$^{\circ} \times$ 0.5$^{\circ}$ grid box. }\label{Fig1:Data_Coverage}
\end{figure}

The primary data used for this study are approximately 250,000 profiles of temperature and salinity from the surface to 2000~db, collected from 1,956 autonomous Argo floats  \citep{RoemmichEtAl2009,RiserEtAl2016}, between 80$^{\circ}$S and 30$^{\circ}$S of latitude, from the 1st of January 2006 to the 31 of December 2014. Argo floats provide broad scale coverage of the Southern Ocean, shown in Fig. \ref{Fig1:Data_Coverage}a, with sufficient spatial and temporal resolution to resolve the large-scale circulation and seasonal variability. Unfortunately, the relatively large distances between observations (approximately 200~km in the Southern Ocean) and the insufficient number of temporally simultaneous measurements means that the Argo array is not capable of directly resolving the instantaneous meso-scale. However, \cite{McCaffreyEtAl2015} have shown that is is possible to use Argo floats to measure the statistics of the meso-scale turbulence. 

%%=========%
%%Figure 1
%%=========%
\begin{figure}[t]
   \centering  
  \includegraphics[width=18pc,height=18pc,angle=0]{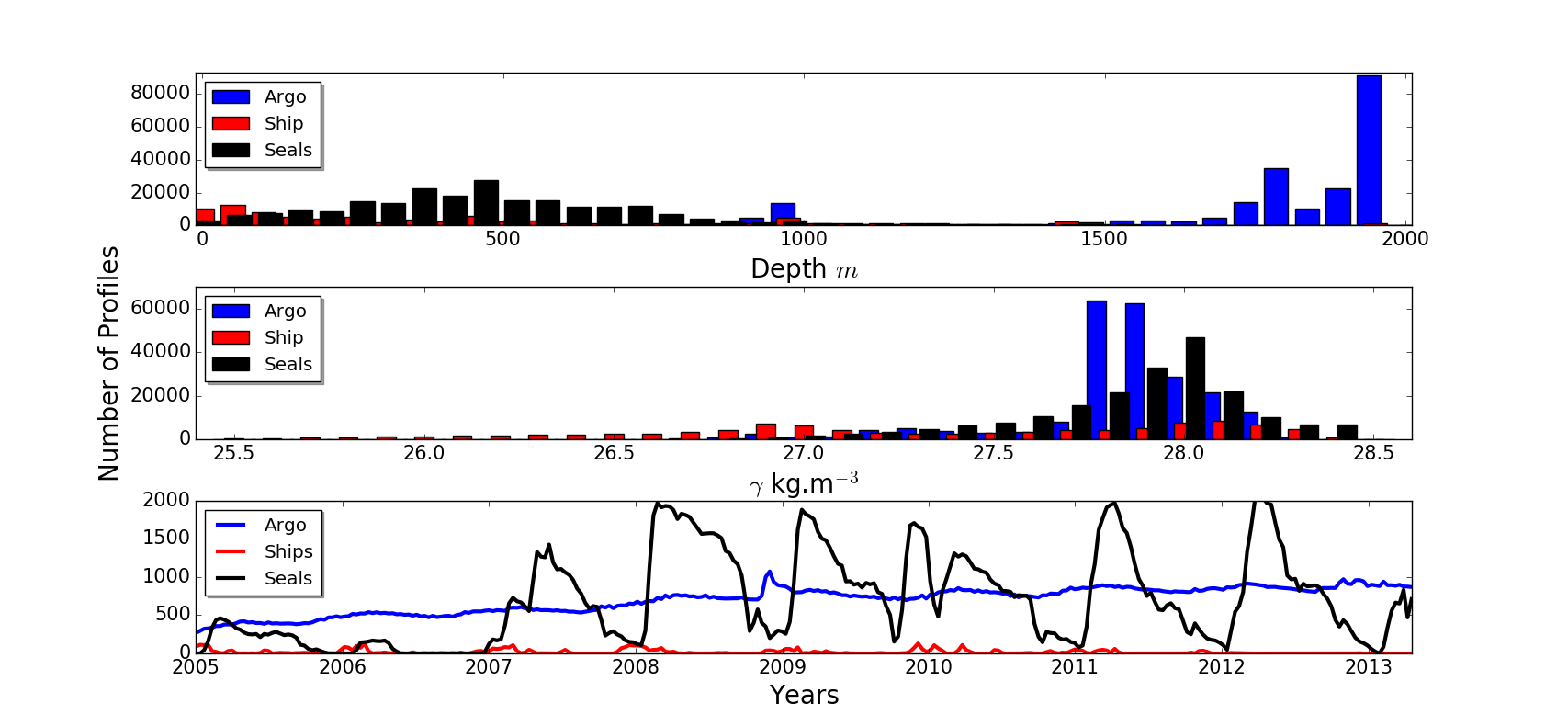}\\
  \caption{Depth, density and temporal sampling of in the hydrographic data in the Southern Ocean. (a) histogram showing the deepest depth sampled by hydrographic profiles sourced from the Argo floats (blue), ship based WOD (red) and instrumented elephant seals (black); (b) as in Fig. \ref{Fig2:Data_Depth_Temporal}a but for the densest neutral density $\gamma$ sampled for each data source; and (c) The number of profiles from each data source, each month, between 2005 and 2014.}\label{Fig2:Data_Depth_Temporal}
\end{figure}

%width=40pc,height=20pc,angle=0
The Argo array has a limited number of observations along the southern border of the ACC, and along the seasonal sea-ice edge. To supplement the Argo data, we additionally employ hydrographic profiles obtained from various research cruises, assembled in the Wold Ocean Database (WOD) \citep{BoyerEtAl2009} and 223,426 profiles collected from 513 instrumented southern elephant seals \citep{RoquetEtAl2013,RoquetEtAl2014}. These animals forage throughout the Southern Ocean, but prioritize regions that are generally further south of those sampled by the Argo floats. Data coverage of Argo, WOD and instrumented seals are shown in \ref{Fig1:Data_Coverage}. The combined dataset samples all of the major water mass classes within the Southern Ocean, as shown in Fig. \ref{Fig2:Data_Depth_Temporal}. This figure shows histograms of the deepest depth (Fig. \ref{Fig2:Data_Depth_Temporal}a) and densest neutral density  (Fig. \ref{Fig2:Data_Depth_Temporal}b) sampled by each of the three data sources. The majority of Argo profiles sample to 2000~m and to about $\gamma$=27.8~kg.m$^{-3}$; the majority of the instrumented seals profiles sample to around 500~m depth, with some profiles deeper than 1000~m, and to about $\gamma$=28.0~kg.m$^{-3}$. We note that although waters denser than $\gamma$=28.0~kg.m$^{-3}$ (Antarctic Bottom Water) are sampled in this dataset, the coverage is patchy, being sampled primarily in the Atlantic Sector. It can be seen in Fig. \ref{Fig2:Data_Depth_Temporal}c that the addition of instrumented seals to the database significantly improves winter data coverage. 

%%=========%
%%Figure 2
%%=========%

For all datasets, only profiles that have passed quality control checks are used. Additional quality control was carried out using automated outlier detection algorithm based on an interquartile range filter and density inversion filter, as in \cite{SchmidtkoEtAl2013}. Data from 2006 to 2014 are used in this study, as there is insufficient data in preceding years to provide coverage of the entire Southern Ocean, as shown in Fig. \ref{Fig2:Data_Depth_Temporal}c.

\subsection{Satellite Data}

In addition to the hydrographic data, we also employ satellite derived estimates of sea-surface dynamic topography in order to provide a ``reference" velocity. Here we use the Archiving, Validation, and Interpretation of Satellite Oceanographic data (AVISO) daily gridded absolute dynamic topography (ADT) from Ssalto/Duacs, downloaded from Copernicus Marine Services (http://marine.copernicus.eu/web/69-interactive-catalogue.php). We use delayed-mode dynamic topography provided on a 1/4$^{\circ}$ Mercator grid, obtained by optimally interpolating the alongtrack data series based on the REF dataset, which uses two satellite missions [Ocean Topography Experiment(TOPEX)/Poseidon/European Remote Sensing Satellite(ERS) or Jason-1/Envisat or Jason-2/Envisat] with consistent sampling over the 21-yr period. 

The AVISO ADT is then calculated at each of the hydrographic profile locations and sampling times by 3-dimensional linear interpolation (that is, spatially and temporally). Thus, for every hydrographic profile we have an associated estimate of the ADT. Profiles obtained in regions or at times where ADT data are not available (which occurs frequently in winter in the far south of the domain) are flagged and excluded from the analysis.

\subsection{Surface Wind Forcing}

To calculate the meridional circulation due to Ekman currents we use the daily mean output of surface momentum flux from the National Center for Environmental Prediction (NCEP)  reanalysis product (http://www.esrl.noaa.gov/psd/data/reanalysis/reanalysis.shtml), described in \cite{KalnayEtAl1996}, to determine the wind stress $\tau$.  

\subsection{Climatology of the Southern Ocean} \label{Sec:Climatology}

Using the hydrographic and satellite data products described above, we develop a climatology of the Southern Ocean. In particular, from the temperature, salinity and pressure profiles we compute the neutral density, $\gamma$, the isopycnal potential vorticity (IPV), $q$, and the absolute geostrophic streamfunction, $M$. 

Profiles of $\gamma$ are computed from our hydrographic database using the software described by \cite{Jackett&McDougall1997}. In order to compute the IPV, we make the planetary geostrophic approximation, which is a good approximation of the Ertel PV in the  Southern Ocean interior \citep{Thompson&NaveriaGarabato2014}:
\begin{equation}   \label{Eqn:PV_definition}
q \approx \frac{f}{\rho_0} \frac{\partial \gamma}{\partial z}.
\end{equation}
To compute profiles of isopycnal streamfunction, defined in \cite{McDougall&Klocker2010} (see Eqn. \ref{Eqn:Geostophic_Streamfunction}), we use version 3 of the TEOS-10 software \citep{McDougall&Barker2011}. From hydrographic data we can only obtain the \textit{relative streamfunction}, that is, the streamfunction relative to some reference level $\gamma=\gamma_{\textrm{ref}}$:
\begin{eqnarray} \label{Eqn:Relative_Streamfunction_Def}
\mathbf{u}_{g_{\textrm{rel}}}(x,y,\gamma) & = & \mathbf{u}_{g}(x,y,\gamma) - \mathbf{u}_{g} (x,y,\gamma_{\textrm{ref}}) \dots \nonumber \\ 
                                          & = & \frac{1}{f}\mathbf{k} \times \nabla M_{\textrm{rel}}.
\end{eqnarray}    
To determine the \textit{absolute} streamfunction we follow \cite{Kosempa&Chambers2014} and reference our streamfunction to the surface. Since the ADT can be interpreted as the surface streamfunction:
\begin{equation} \label{Eqn:Surface_Streamfunction_Def} 
\mathbf{u}_{\textrm{g}}(x,y,\gamma_{\textrm{surf}}) = \frac{g}{f} \mathbf{k} \times \nabla ADT,
\end{equation}
the absolute streamfunction is computed by adding the estimated ADT at each hydrographic profile location to the relative streamfunction referenced to the surface:
\begin{equation} \label{Eqn:Absolute_Streamfunction_Def} 
M_{\textrm{abs}} = ADT+M_{\textrm{rel}}.
\end{equation}

Finally, profiles of neutral density, IPV and absolute geostrophic streamfunction are interpolated to a regular longitude/latitude grid using the CARS--LOWESS (CSIRO Atlas of Regional Seas robust LOcally Weighted regrESSion) software \citep{RidgwayEtAl2002}. The neutral density is mapped on depth surfaces from the surface to 2000~m, with a vertical spacing of $\Delta z$=50~m.  The IPV and streamfunction are mapped on a set of isopycnal layers from $\gamma$=26.0~kg.m$^{-3}$ to $\gamma$=28.5~kg.m$^{-3}$, with a vertical spacing of $\Delta \gamma$=0.05~kg.m$^{-3}$. For consistency with the altimetric observations, we use a horizontal grid spacing of 0.25$^{\circ}\times$0.25$^{\circ}$, although the effective resolution of the hydrographic data is coarser  (the average distance between Argo floats profile locations in the Southern Ocean is approximately 200~km). 

\begin{figure*}[t]
   \centering  
  \includegraphics[width=40pc,height=20pc,angle=0]{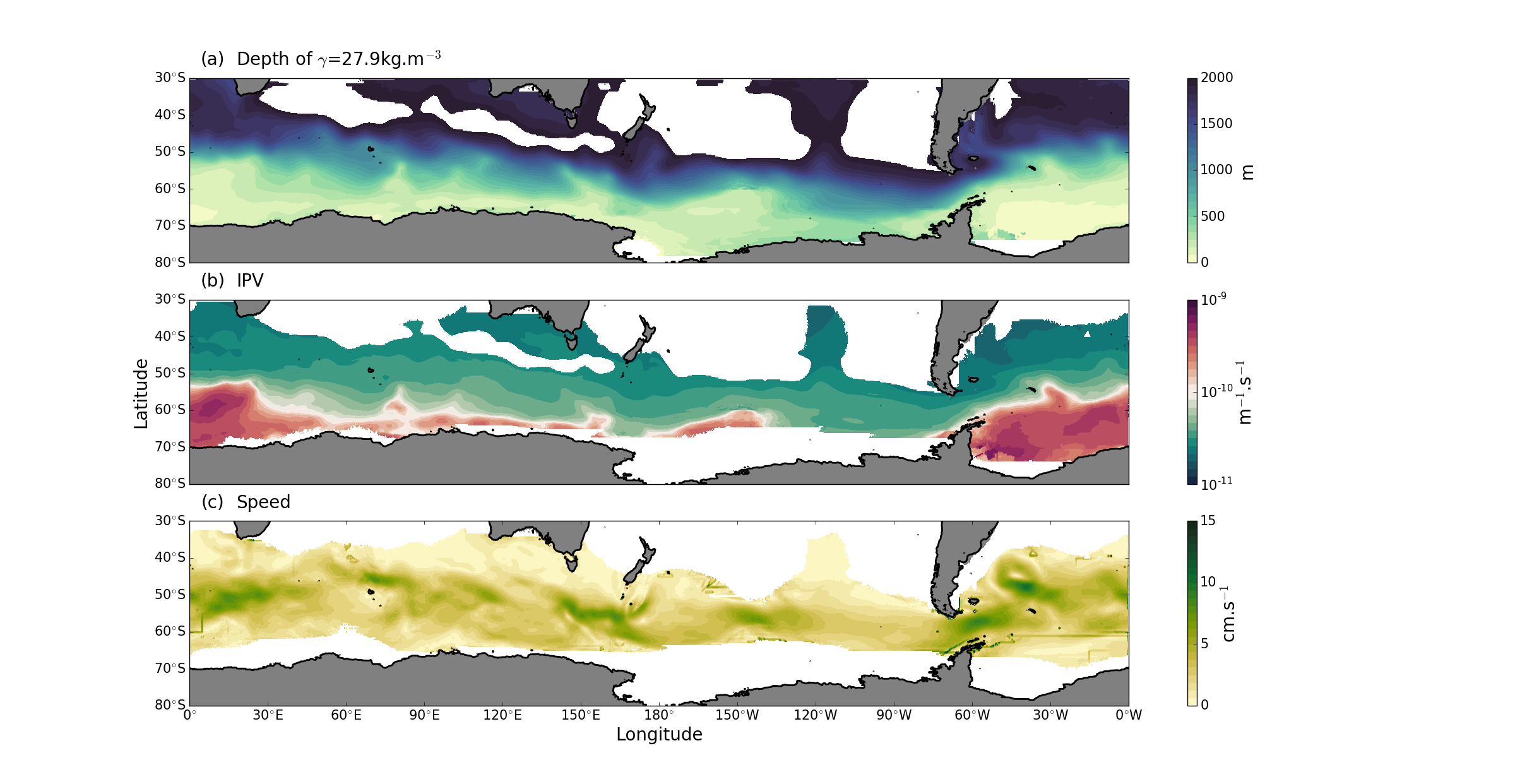}\\
    \caption{The climatology of the Southern Ocean, determined from combined hydrographic/satellite data set, on isopycnal $\gamma$=27.9~kg.m$^{-3}$. (a) The depth of the isopycnal; (b) the isopycnal potential vorticity (IPV); and (c) the zonal current speed. Blanked out areas indicate regions with fewer than 200 data points in the interpolation process, or, where the isopycnal is deeper than 2000m. Note the logarithmic colorscale used for the IPV in Fig. \ref{Fig3:PV_gradPV_U_zonal_ave}b}\label{Fig3:PV_gradPV_U_zonal_ave}
\end{figure*}

An example of our climatology is shown in Fig. \ref{Fig3:PV_gradPV_U_zonal_ave}, here for the isopycnal $\gamma=$27.9~kg.m$^{-3}$. The depth of this isopycnal is shown in Fig. \ref{Fig3:PV_gradPV_U_zonal_ave}a, which reveals, as expected, isopycnals shoaling towards higher latitudes and eventually outcropping with the surface near the Antarctic continent. We note that although this isopycnal is well represented in our dataset (see Fig. \ref{Fig2:Data_Depth_Temporal}b), it is deeper than 2000~m over much of region north of the ACC. Fig. \ref{Fig3:PV_gradPV_U_zonal_ave}b shows IPV on the same isopycnal, which increases poleward, as expected. However, it is worth noting that the IPV structure is not zonally homogeneous, and there are regions of stronger and weaker meridional gradients, which according to Eqn. \ref{Eqn:PV_Flux}, can indicate regions of enhanced eddy volume transport. Finally, Fig. \ref{Fig3:PV_gradPV_U_zonal_ave}c shows the geostrophic current speed computed from the gradient of the absolute geostrophic streamfunction gradient. The currents appear realistic: they form jets, and show steering by topography. The strength of these mean currents is important for the suppression of eddy volume fluxes, as will become apparent in sections \ref{Section:K_Calculation} and \ref{Section:Reconstruction_Results}.   

%%=========%
%%Figure 3
%%=========%

%width=40pc,height=20pc,angle=0

%Data are corrected for instrumental errors, atmospheric perturbations, orbit errors, tides, inverted barometer bias, and aliased fast barotropic signals (periods of less than 20 days). 

%Satellite altimeters are only capable of measuring anomalous sea-surface height. To estimate the mean state of the ocean we use the combined mean dynamic topography (MDT) product, described in \cite{RioEtAl2014} and downloaded from CNES-CLS13 (http://www.aviso.altimetry.fr/fr/donnees/produits/produits-auxiliaires/mdt.html). We then form the ``absolute dynamic topography" (ADT) which is simply the sum of the time-mean and time-varying sea level. 

%The mean sea surface is reconstructed over the period 1993--2012 by combining data from the Gravity Recovery and Climate Experiment (GRACE) mission, satellite altimeters, and drifting buoys. 

\section{The Three-Dimensional Eddy Diffusivity} \label{Section:K_Calculation}
In this section, we use the hydrographic profiles to determine a three-dimensional estimate of both the suppressed $\tilde{K}$ and unsuppressed $K_0$ eddy diffusivity, following the theoretical framework described by \cite{Ferrari&Nikurashin2010}, described in Sec. \ref{Section:MOC_Theory}. Recall Eqn. \ref{Eqn:Eddy_Diffusivity}, in which the total eddy diffusion is written as an unsuppressed diffusivity $K_0$ multiplied by a suppression factor that describes the influence of the mean flow on the eddy stirring. 

\begin{figure*}[t]
   \centering  
  \includegraphics[width=40pc,height=20pc,angle=0]{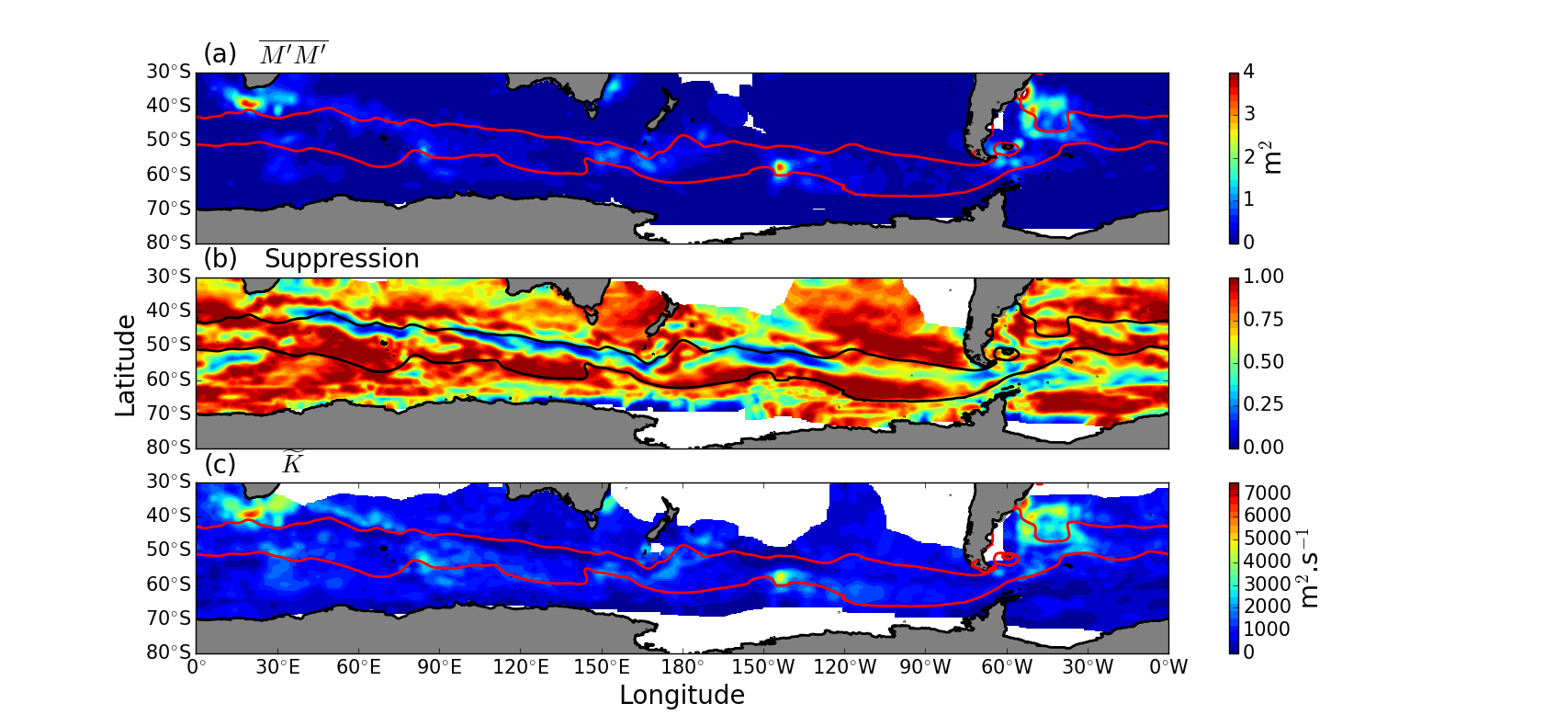}\\
    \caption{Meridional isopycnal diffusivity in the Southern Ocean on isopycnal $\gamma$=27.9~kg.m$^{-3}$: (a) the variance of the isopycnal geostrophic streamfunction $\overline{M^{\prime}M^{\prime}}$; (b) the diffusivity suppression factor $1/\left [1+k^{2}_\textrm{eddy}\tau_{\textrm{eddy}}^{2}\left(c_p -\overline{u}\right)^{2}\right ]$; and (c) the suppressed eddy diffusivity $\tilde{K}$. Solid lines indicate the positions of the northern and southern ACC boundaries. }\label{Fig4:Diffusivity_Map_DIMES_isopycnal}
\end{figure*}

In order to compute the unsuppressed diffusivity, we use the expression introduced by \cite{Holloway1986} and \cite{Keffer&Holloway1988} that relates the root-mean-square of the streamfunction fluctuations to $K_{0}$:
\begin{equation}
K_{0} = \frac{\Gamma}{f} \left( \overline{M^{\prime} M^{\prime}} \right)^{1/2}
\end{equation}
where $\Gamma$ is a constant \textit{mixing efficiency}, usually taken to be 0.35 \citep{Klocker&Abernathey2014}. We compute the RMS of the geostrophic streamfunction by first computing the streamfunction fluctuations by subtracting the mean geostrophic streamfunction, $\overline{M}$, from each of the instantaneous profiles of $M$. The square of $M^{\prime}$ is then computed for each profiles, and mapped using the CARS--LOWESS software on a regular longitude/latitude grid (see Fig. \ref{Fig4:Diffusivity_Map_DIMES_isopycnal}a). A highly zonally assymetric field is produced, with elevated streamfunction variance found in regions downstream of large bathymetric features, in western boundary currents, in the Agulhas region ($\sim$ 20\---60$^\circ$E), and at the central Pacific Fracture Zone ($\sim$ 140$^\circ$W), consistent with previous studies \citep{SalleeEtAl2008,Klocker&Abernathey2014,RoachEtAl2016}. 

To compute the suppression factor, we require an estimate of the time-mean current velocity, $\overline{u}$, the eddy phase speed, $c_p$, the eddy decorrelation time-scale, $\tau_{\textrm{eddy}}$, and the eddy wavenumber, $k_{\textrm{eddy}}$. $\overline{u}$ is obtained from the absolute geostrophic streamfunction $\overline{M}$, as described in Sec. \ref{Section:Data_Methods}\ref{Sec:Climatology}. $c_p$ is calculated using Rossby wave dispersion relationship, Doppler shifted by the depth mean flow, as suggested by \cite{Klocker&Marshall2014}:
\begin{equation} \label{eddy_speed}
c_p = \overline{u}^{zt} - \beta L_{D}^{2},
\end{equation}
where $\beta$ is the meridional gradient of the Coriolis parameter, $\overline{u}^{zt}$ is the depth averaged zonal velocity and $L_D$ is the first baroclinic deformation radius. To compute $L_D$, we solve the Sturm-Liouville problem for the neutral-modes of the linearized quasi-geostrophic equation using the finite difference scheme of \cite{Smith2007} and our gridded interpolated neutral density. The maps of $L_D$ (not shown) produced by this calculation are very similar to those of \cite{CheltonEtAl1998}, although  due to the more complete data coverage provided by the Argo floats, there are fewer regions with missing data and we find a larger deviation of contours of constant $L_D$ near large bathymetric features. The eddy decorrelation time-scale,  $\tau_{\textrm{eddy}}$ is taken to be a constant 4 days, as found by \cite{Klocker&Abernathey2014}. Finally, the eddy length scale, used in the calculation of eddy wavenumber $k_{\textrm{eddy}}=2 \pi/L_{\textrm{eddy}}$, is estimated by assuming a constant ratio between the eddy size and $L_D$, which is approximately valid for strongly non-linear eddies, such as those found in the Southern Ocean \citep{Klocker&Abernathey2014}. We set this ratio to 2.5, so that $L_{\textrm{eddy}}=$2.5$L_{D}$.

% * <jean-baptiste.sallee@locean-ipsl.upmc.fr> 2016-10-26T10:32:26.037Z:
% 
% > $\tau_{\textrm{eddy}}$ is taken to be a constant 4 days, as found by \cite{Klocker&Abernathey2014}. Finally, the eddy length scale, used in the calculation of eddy wavenumber $k_{\textrm{eddy}}=2 \pi/L_{\textrm{eddy}}$, is estimated by assuming a constant ratio between the eddy size and $L_D$, which is approximately valid for strongly non-linear eddies, such as those found in the Southern Ocean \citep{Klocker&Abernathey2014}. We set this ratio to 2.5, so that $L_{\textrm{eddy}}=$2.5$L_{D}$. 
% I think these choices are discutable. But if we are questioned on those, the point is that they indeed might change intensity of suppresion etc., of K_tilde, and therefore resulting overturning, but they are realistic and allow to test our hypothesis that realistic 3D variations and (ACC-induced) suppression of kappa do impact overturning.
% 
% ^.

With all the ingredients assembled, we compute the suppression factor, which is plotted on isopycnal $\gamma$=27.9~kg.m$^{-3}$ in Fig. \ref{Fig4:Diffusivity_Map_DIMES_isopycnal}b. Several regions of heavily suppressed diffusivities (with a suppression factor between 0 and 0.25) are found in regions of strong zonal jets (compare with Fig. \ref{Fig3:PV_gradPV_U_zonal_ave}c), as expected. The suppression factor computed here shows a strong qualitative resemblance to those computed by \cite{Ferrari&Nikurashin2010} and \cite{Klocker&Abernathey2014} at the surface using altimetry alone. 

%%=========%
%%Figure 4
%%=========%

The geographical distribution of the suppressed eddy diffusivity, once again for the isopycnal $\gamma$=27.9~kg.m$^{-3}$, is plotted in Fig. \ref{Fig4:Diffusivity_Map_DIMES_isopycnal}c. Our map appears realistic and shows similar features to the estimate of $\widetilde{K}$ by \cite{ColeEtAl2015} using estimate of the mixing length obtained by considering the decorrelation length-scale of salinity fluctuations measured by Argo floats. In particular, we note enhanced regions of $\widetilde{K}$ downstream of large topographic features where both streamfunction fluctuations are strong and time-mean flows are weak. When zonally integrated and mapped back to depth coordinates, as shown in Fig. \ref{Fig5:Diffusivity_With_Depth}, we see that the unsuppressed diffusivity $K_0$ is strong at the surface and decreases with depth (Fig. \ref{Fig5:Diffusivity_With_Depth}a). In contrast, $\widetilde{K}$ is enhanced at depth, reaching a peak at about 1000m. This peak in $\widetilde{K}$ is found very close to the steering level (where $cp \approx \overline{u}$) predicted by \cite{Smith&Marshall2009} using linear theory, and that observed by \cite{ColeEtAl2015}, although it is shallower than steering level found in \cite{AbernatheyEtAl2010}'s eddy permitting simulation (found at about 1750m). 

\begin{figure}[t]
   \centering  
  \includegraphics[width=18pc,height=20pc,angle=0]{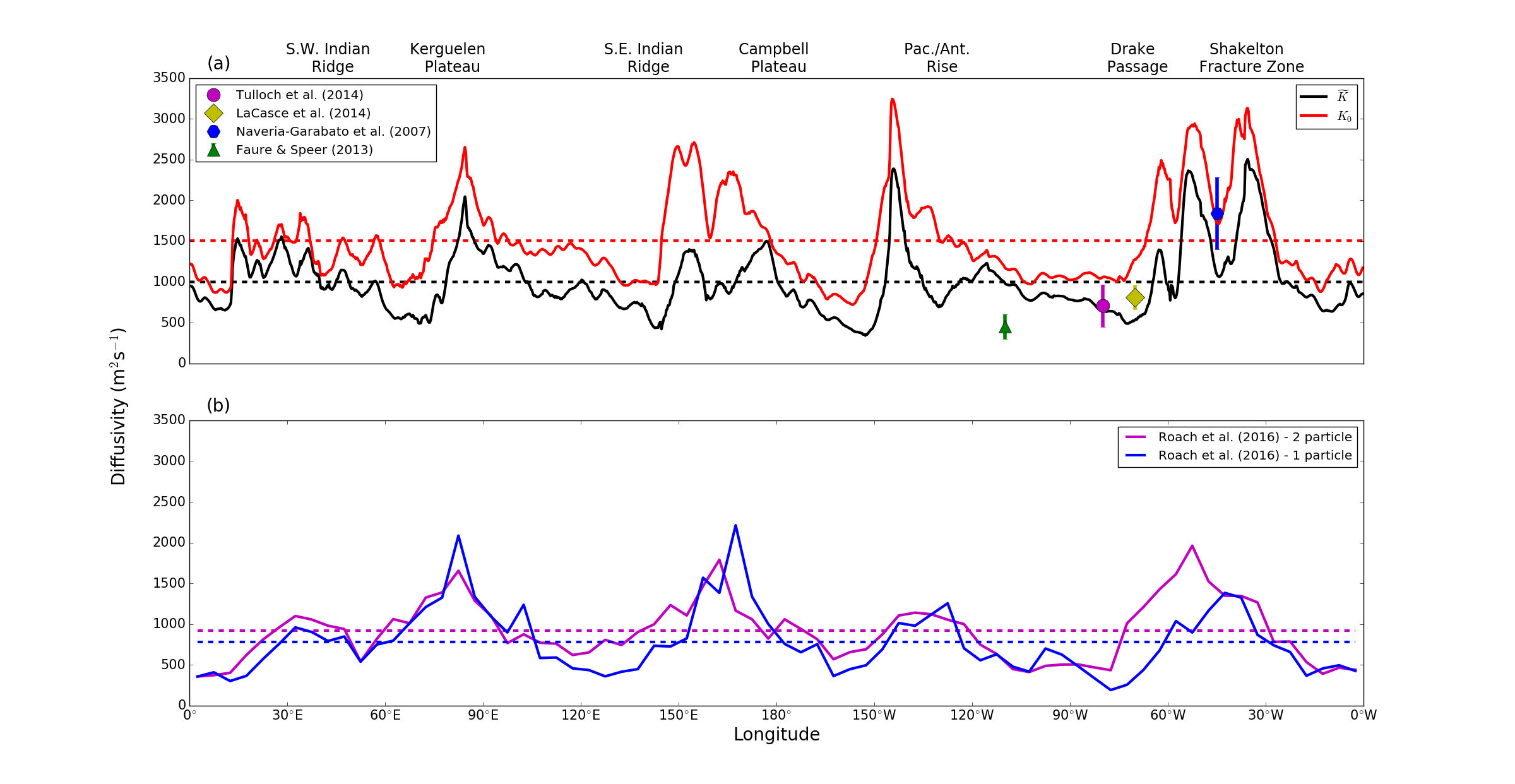}\\
  \caption{Variation of diffusivity with longitude: (a) The unsupressed diffusivity $K_0$ (red) and the supressed diffusivity $\tilde{K}$ (black) averaged over the ACC envelope (solid lines) and their zonal average (dashed lines) on isopycnal $\gamma=27.9$~kg.m$^{-3}$. For comparison, the estimates of the diffusivity from previous studies are shown as solid markers; and (b) the one and two particle estimates of the diffusivity from \cite{RoachEtAl2016} at 1000~m depth, included for comparison.}\label{Fig6:Diffusivity_With_Longitude}
\end{figure}

To underscore the important role that bottom bathymetry plays in controlling the diffusivity, we plot $K_0$ (red) and $\widetilde{K}$ (black) on the isopycnal $\gamma$=27.9~kg.m$^{-3}$ in Fig. \ref{Fig6:Diffusivity_With_Longitude}a, but now meridionally averaged from the southern boundary of the ACC to the northern boundary of the ACC (determined by finding contours of MDT that correspond to the Southern ACC Front and the Subantarctic Front, as in \cite{Sokolov&Rintoul2007}, plotted as solid lines in Fig. \ref{Fig4:Diffusivity_Map_DIMES_isopycnal}). The zonal mean of the $K_0$ and $\widetilde{K}$ (dashed lines in Fig. \ref{Fig6:Diffusivity_With_Longitude}) show that the suppressing effect of the mean-flow acts to reduce the diffusivity by about 500~m$^{2}$s$^{-1}$ in the ACC latitudes. However the suppressed diffusivity still peaks downstream of large bathymetric features, reaching its maximum values downstream of the Pacific Antarctic Rise ($\sim$140$^{\circ}$W), and downstream of Drake Passage ($\sim$60$^{\circ}$W; Fig. \ref{Fig6:Diffusivity_With_Longitude}). The suppressing effect of the mean-flow is perhaps most clearly seen at Southeastern Indian Ridge and the Campbell Plateau ($\sim$150\---170$^{\circ}$E), where $K_0$ peaks at about 2500~m$^{2}$s$^{-1}$, but the suppressed diffusivity does not rise above 1500~m$^{2}$s$^{-1}$; a local effective suppression of about 1000~m$^{2}$s$^{-1}$. The spatial structure of our estimate is similar to that of \cite{RoachEtAl2016} shown in Fig. \ref{Fig6:Diffusivity_With_Longitude}b, who used the dispersion of Argo floats at 1000~m to directly estimate cross-stream diffusivity. \cite{RoachEtAl2016}'s estimate shows peaks in similar locations to ours, with similar magnitudes, although our estimates of $\widetilde{K}$ are substantially lower than theirs at the Campbell Plateau once the suppression factor is applied. The differences between our estimates may arise due to the different formulation used in the estimate, but more likely due to the fact that the \cite{RoachEtAl2016} estimate was made at 1000~m, whereas the isopycnal $\gamma$=27.9~kg.m$^{-3}$ is closer to 1500~m in the ACC (see Fig. \ref{Fig3:PV_gradPV_U_zonal_ave}a).    

%%=========%
%%Figure 6
%%=========%

We compare our estimated diffusivities with several estimates made near Drake Passage by direct measurement \citep{NaveiraGarabatoEtAl2007,Faure&Speer2012,LaCasceEtAl2014,TullochEtAl2014}. We find that our estimate of the effective diffusivity agrees broadly with these other estimates, although we note that our estimates are significantly higher than those of \cite{Faure&Speer2012} and somewhat lower than those of \cite{NaveiraGarabatoEtAl2007}. However, given the difficulty in estimating certain parameters in the suppression factor, the reasonably close agreement between our estimate of $\widetilde{K}$ and previous local or regional estimates gives us some confidence in our maps of eddy diffusivity.
 
\section{Reconstruction of the MOC} \label{Section:Reconstruction_Results}

As described in Section \ref{Section:MOC_Theory} the MOC can be decomposed into an time-mean Ekman component $\overline{h}v_{\textrm{Ekman}}$, a time-mean geostrophic component $\overline{h}v_{\textrm{g}}$, and the transient eddy component  $\overline{h^{\prime}v_{\textrm{g}}^{\prime}}$. In this section, we compute each of these components from observations in order to reconstruct the residual overturning streamfunction and described how it is influenced by the spatial variation and suppression of the diffusivity. 

\subsection{Eulerian Mean Overturning}

\begin{figure}[t]
   \centering  
  \includegraphics[width=18pc,height=20pc,angle=0]{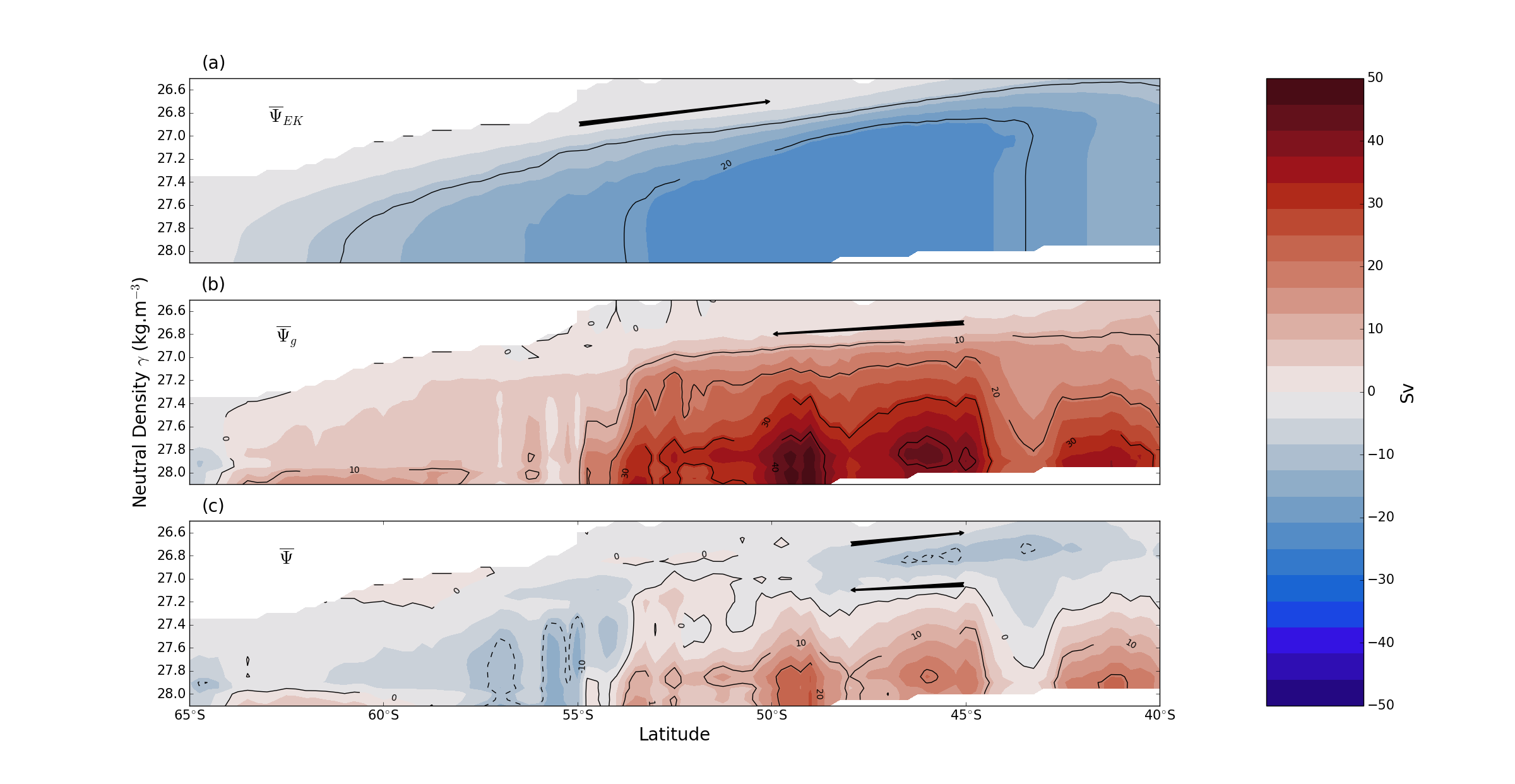}\\
  \caption{The time-mean overturning streamfunction: (a) the Ekman overturning $\overline{\Psi}_{\textrm{Ekman}}$; (b) the geostrophic overturning $\overline{\Psi}_{\textrm{g}}$; and (c) the total time-mean overturning $\overline{\Psi}=\overline{\Psi}_{\textrm{Ekman}}+\overline{\Psi}_{\textrm{g}}$. Positive (negative) values indicate clockwise (counter-clockwise) transport, as indicated by the arrows.   }\label{Fig7:Time_Mean_Overturning}
\end{figure}

The components of the time-mean overturning are determined by computing the Ekman ageostrophic velocity from the equations for an Ekman spiral (Eqns. \ref{Eqn:Ekman_0} and \ref{Eqn:Ekman_1}), and by computing the time-mean geostrophic velocity $\mathbf{u}_{\textrm{g}}$ from the absolute geostrophic streamfunction, $M$ and Eqns. \ref{Eqn:Geostophic_Streamfunction}. We then determine a time-mean isopycnal layer thickness $\overline{h}$ by simply taking the difference in the depths of the isopycnal layer interfaces:
\begin{equation}
\overline{h}_{j} = z_{j+1/2}-z_{j-1/2} 
\end{equation}
where $j$ is the index of the $j$th isopycnal layer. Finally, the results are integrated zonally and vertically to give the mean overturning streamfunctions, $\overline{\Psi}_{\textrm{Ekman}}$, $\overline{\Psi}_{\textrm{g}}$ and the total mean overturning $\overline{\Psi} = \overline{\Psi}_{\textrm{Ekman}} + \overline{\Psi}_{\textrm{g}}$. These streamfunctions are plotted in Fig. \ref{Fig7:Time_Mean_Overturning}.

The zonally integrated Ekman driven overturning  $\overline{\Psi}_{\textrm{Ekman}}$, shown in Fig.  \ref{Fig7:Time_Mean_Overturning}a, consists of a single clockwise overturning cell that transports around 20~Sv of water northwards at the surface, and drives a strong upwelling between $65^{\circ}$S and $55^{\circ}$S, which corresponds to the unblocked latitudes of Drake Passage. The Ekman circulation is largely opposed by the mean geostrophic overturning, $\overline{\Psi}_{\textrm{g}}$ (Fig. \ref{Fig7:Time_Mean_Overturning}b), that consists of a counter-clockwise overturning cell with a peak transport of more than 40~Sv near $\gamma=$28.0~kg.m$^{-3}$. The geostrophic overturning cell additionally drives a weak downwelling in the Drake Passage latitudes. Both the geostrophic and Ekman overturning cells show strong similarity to those obtained from the data-assimilating Southern Ocean State Estimate (SOSE) model \citep{Mazloff2008,Mazloff2013} (in particular, see Fig. 4-5 of \cite{Mazloff2008}). The concordance between SOSE and our estimates is perhaps not surprising, given that SOSE assimilates both Argo hydrographic data and satellite altimetry. However, comparing our results to the SOSE output does give some confidence that the analysis of the hydrographic profiles has been conducted correctly and that the absolute streamfunction computed from the combined altimetry/hydrography is giving realistic results.

The total mean overturning $\overline{\Psi}$, (Fig. \ref{Fig7:Time_Mean_Overturning}c) shows the effect of compensation of the Ekman driven overturning by the geostrophic overturning. As noted by \cite{Mazloff2008} and \cite{Mazloff2013}, in much of the region north of Drake Passage (i.e. north of 55$^{\circ}$S), the geostrophic component of the overturning dominates the Ekman transport, leading to a net southward transport of Circumpolar Deep Waters (those waters denser than about 27.5~kg.m$^{-3}$), although it must be emphasized that within Drake Passage latitudes, the Ekman driven upwelling still dominates the geostrophic downwelling, and that the northward transport due to the Ekman currents remains dominant near the surface in the lighter water classes. Our results strongly echo those of \cite{Mazloff2008}, and stress the importance of the interior geostrophic component for the overturning. We note that this component is often ignored in analyses of the overturning, and is not well incorporated into TEM theories of the overturning.

\subsection{Eddy Overturning}

We now discuss the contributions of transient geostrophic eddies to the MOC. Here we employ the simple downgradient diffusive closure given by Eqn. \ref{Eqn:Geo_Eddy_Vol_flux_2}. To understand the influence of the suppression of the eddy diffusivity by the mean flow on the overturning, we reconstruct the eddy volume flux using both the unsuppressed diffusivity $K_{0}$, and the suppressed diffusivity $\widetilde{K}$. 

\begin{figure*}[t]
   \centering  
  \includegraphics[width=40pc,height=20pc,angle=0]{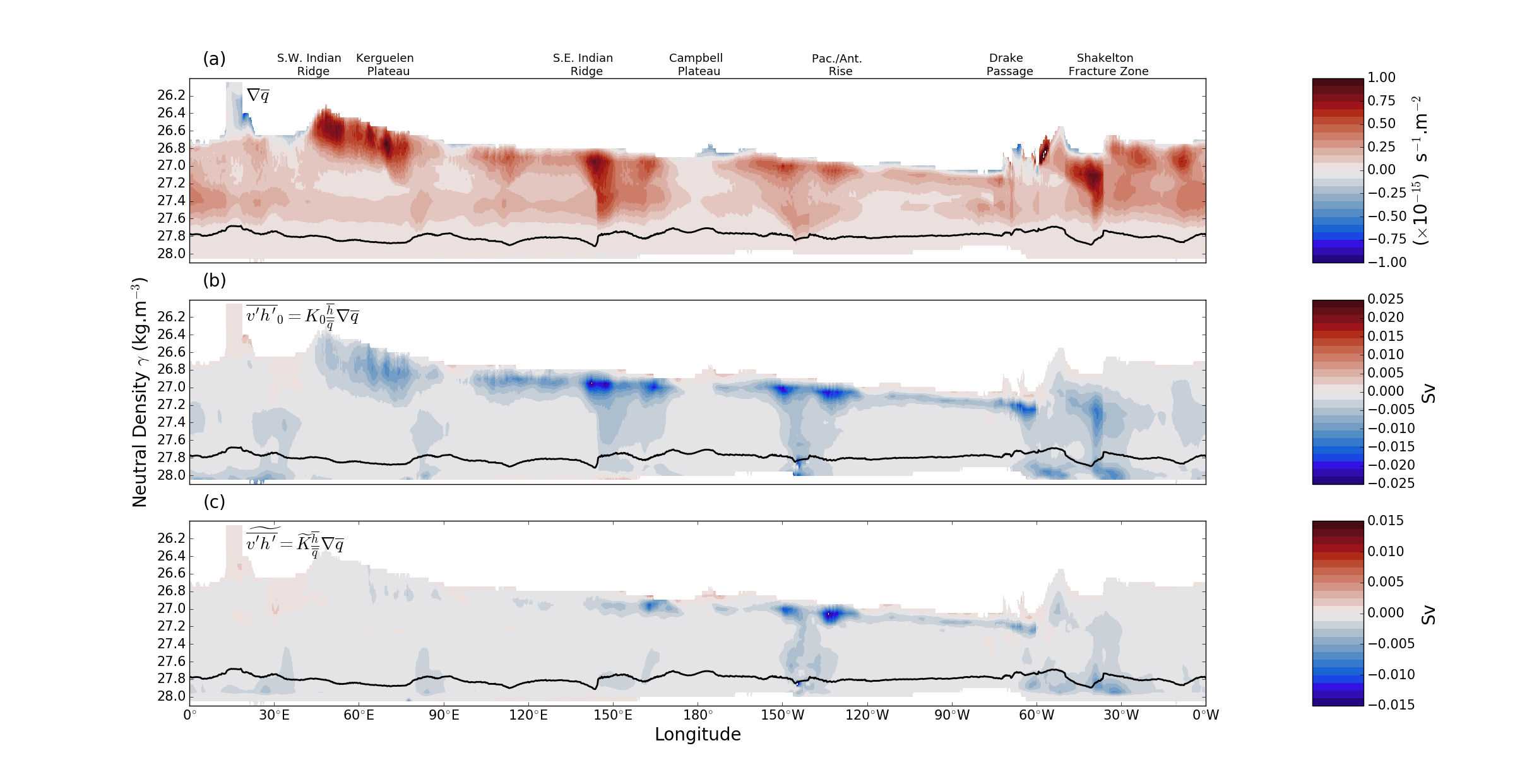}\\
  \caption{The relationship between the meridional IPV gradient and the parameterized eddy flux averaged over the ACC latitude envelope: (a) The meridional IPV gradient; (b) the derived geostrophic eddy-volume flux $\overline{h^{\prime}v^{\prime}}$, computed from Eqn. \ref{Eqn:Geo_Eddy_Vol_flux_2} using the unsuppressed eddy diffusivity $K_{0}$; and (c) as in \ref{Fig8:PV_grad_Eddy_flux}b, using the effective eddy diffusivity $\tilde{K}$. Solid black lines denote the average 1000m depth contour, the approximate critical layer depth. Note the differing colorscales between the unsuppressed (panel b) and suppressed (panel c) transport. }\label{Fig8:PV_grad_Eddy_flux}
\end{figure*}

The longitudinal/vertical structure of meridional IPV gradient, and its relationship with the parameterized eddy fluxes is plotted in Fig. \ref{Fig8:PV_grad_Eddy_flux}, which shows the meridional IPV gradient (Fig. \ref{Fig8:PV_grad_Eddy_flux}a), and estimates of the eddy volume flux using both suppressed and unsuppressed diffusivities (Fig. \ref{Fig8:PV_grad_Eddy_flux}b,c), averaged over the ACC envelope. Despite the argument that IPV should be relatively homogenized in the ocean interior \citep{MarshallEtAl93}, we find substantial IPV gradients in certain regions, particularly downstream of large bathymetric features, a fact that has been remarked upon by previous authors \citep{Thompson&NaveriaGarabato2014}. As a result, both the unsuppressed and suppressed eddy fluxes (Fig. \ref{Fig8:PV_grad_Eddy_flux}b,c) are concentrated in regions donwstream of topography, which is also consistent with previous work \citep{Thompson&Sallee2012,DufourEtAl2015,Chapman&Sallee2016}. Additionally, we note that there is a change in the sign of the IPV gradient in the lighter, surface waters, leading to a \textit{northward} volume transport near the surface, in contrast to the \textit{southward} eddy transport in the interior. The northward eddy flux of light waters, consistent with the negative near-surface IPV gradients, was discussed in depth by \cite{Mazloff2008}.

%%=========%
%%Figure 8
%%=========%

The suppression of the eddy-flux by the mean-flow can be seen by comparing the transports computed with the unsuppressed (Fig. \ref{Fig8:PV_grad_Eddy_flux}b) and suppressed (Fig. \ref{Fig8:PV_grad_Eddy_flux}c) diffusivities. As expected, the eddy volume transports are much larger when the unsuppressed diffusivity is used in the reconstruction. This is particularly evident in the region near the base of the thermocline where the IPV gradient changes sign. When mean-flow suppression is taken into account, the majority of the near surface transport disappears. Additionally, the vertical structure of the transport varies between the suppressed and unsuppressed cases. The unsuppressed transport shows a vaguely equivalent barotropic structure, while the suppressed eddy transport shows minimal interior transports away from the Pacific-Antarctic Rise (between 150$^{\circ}$W and 130$^{\circ}$W) and Drake Passage (between 40$^{\circ}$W and 30$^{\circ}$W). While the unsuppressed transport is typically strongest near the surface, the interior suppressed transport is intensified near the critical layer (at approximately 1000~m depth, indicated by the solid black line in Fig. \ref{Fig7:Time_Mean_Overturning}). Deep transports are, in general, southward.

\begin{figure}[t]
   \centering  
  \includegraphics[width=18pc,height=20pc,angle=0]{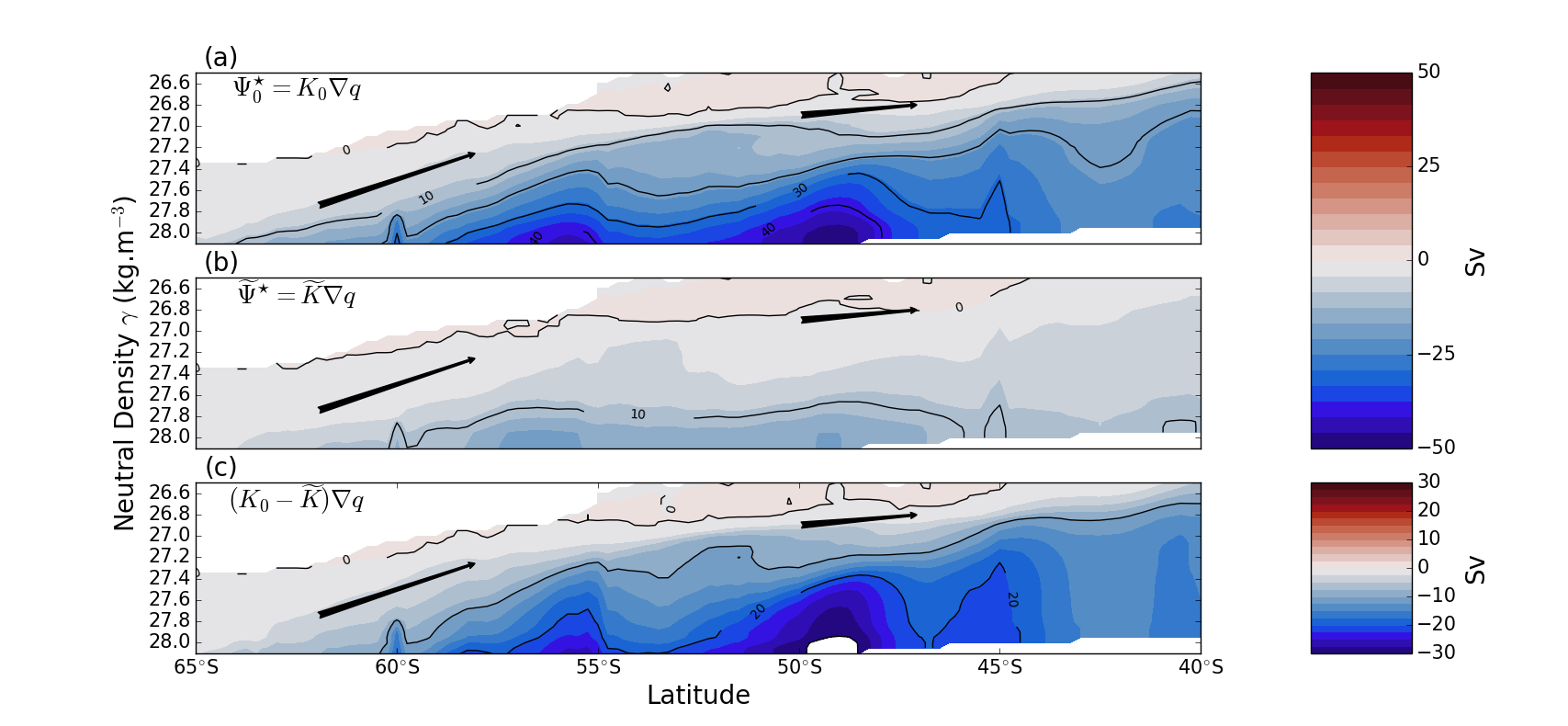}\\
  \caption{The geostrophic eddy overturning streamfunction computed from Eqn. \ref{Eqn:PV_Flux}. (a) $\Psi^{\star}$ computed using the unsuppressed diffusivity $K_0$; (b)  $\Psi^{\star}$ computed using the suppressed diffusivity $\tilde{K}$ (c) The difference between the eddy overturning streamfunctions computed using $K_0$ and $\tilde{K}$. Positive (negative) streamfunction values denote counter-clockwise (clockwise) flow - as indicated by the black arrows. Note the different colorscale used for panel (c).  }\label{Fig9:Eddy_Overturning_Decomposed}
\end{figure}

In Fig. \ref{Fig9:Eddy_Overturning_Decomposed}, we plot the parameterized zonally integrated eddy overturning streamfunction computed using the unsuppressed (Fig. \ref{Fig9:Eddy_Overturning_Decomposed}a) and the suppressed (Fig. \ref{Fig9:Eddy_Overturning_Decomposed}b) diffusivities, as well as the difference between them (Fig. \ref{Fig9:Eddy_Overturning_Decomposed}c). We note that although the parameterization used here is extremely crude, we are able to capture a surprisingly large degree of the eddy-overturning streamfunction computed from the eddy-permitting SOSE model \citep{Mazloff2008,Mazloff2013}. In particular, the overturning streamfunction is generally clockwise in a latitude-density plane, for both suppressed and unsuppressed diffusivities. We note a weak northward flow in the light, near-surface waters that generally reinforce the Ekman currents, with upwelling in the Drake Passage latitudes. In contrast to the SOSE output, our calculations show a general increase in the strength of the eddy overturning streamfunction with depth, although this feature is not as strong (i.e. more consistent with SOSE) when the diffusivity is suppressed. We find southward overturning transports of around 10~Sv at $\gamma=27.0$~kg.m$^{-3}$ at 55$^{\circ}$S when computed with suppressed diffusivities, increasing to around 45~Sv at $\gamma=28.0$~kg.m$^{-3}$ when using the unsuppressed diffusivity. These transport are different by about a factor two for the suppressed diffusivity case ($K=\widetilde{K}$): around 5~Sv at $\gamma=27.0$~kg.m$^{-3}$ at 55$^{\circ}$S, increasing to around 20~Sv at $\gamma=28.0$~kg.m$^{-3}$. For comparison, \cite{Mazloff2008} finds maximum eddy overturning transport of between 10 and 25~Sv, depending on the season. 

%%=========%
%%Figure 9
%%=========%

\begin{figure}[ht]
   \centering  
  \includegraphics[width=18pc,height=18pc,angle=0]{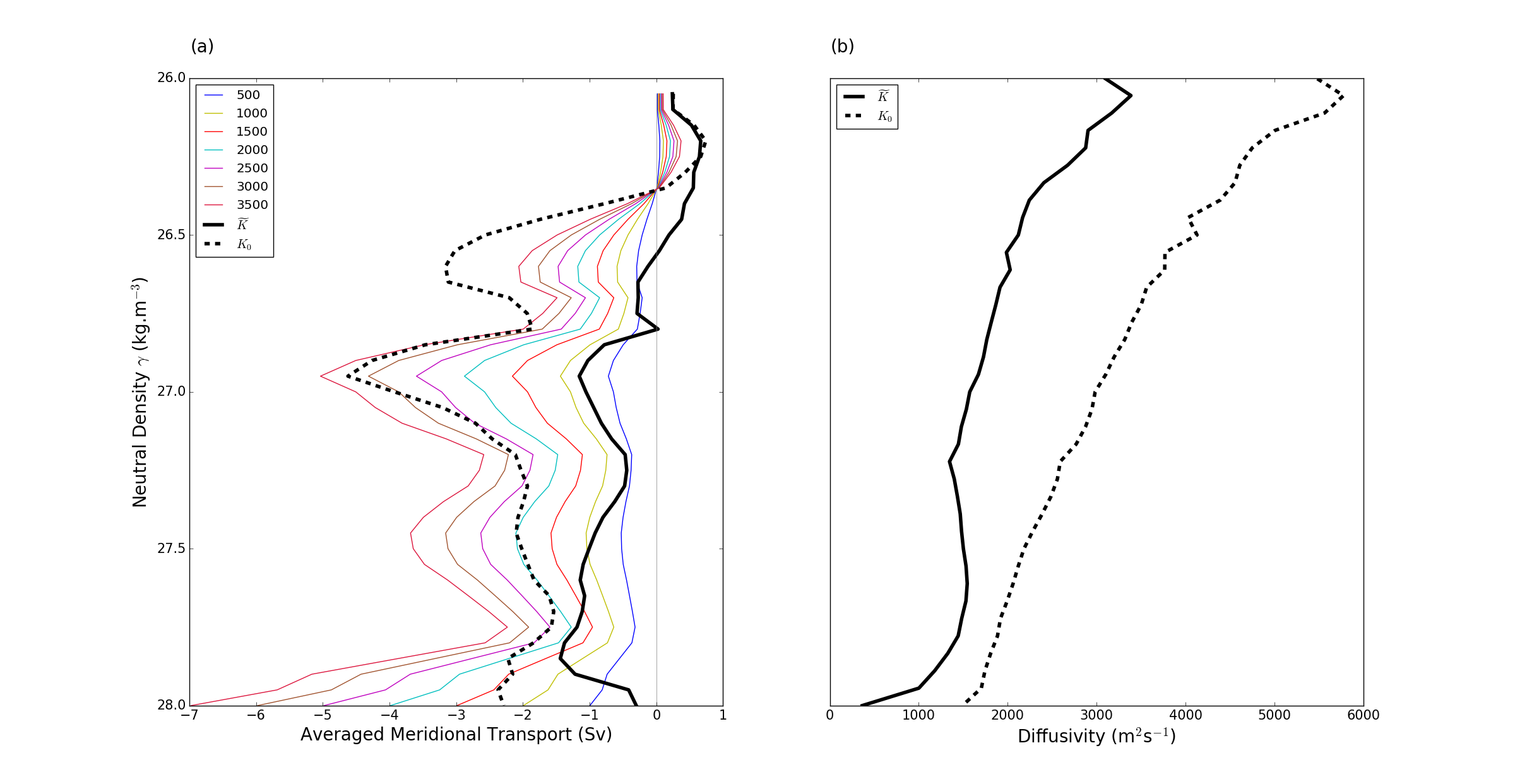}\\
  \caption{The meridional geostrophic eddy transport $\overline{v^{\prime}h^{\prime}}=K\frac{\overline{h}}{\overline{q}}\partial \overline{q}/\partial y$, using variable and constant diffusivities. (a) the meridional transport, zonally integrated and averaged over the ACC latitudes, obtained from Eqn. \ref{Eqn:Geo_Eddy_Vol_flux_2} using the suppressed and unsuppressed spatial variable diffusivity (thick black lines, $K=K_0$ dashed, $K=\widetilde{K}$ solid) and using a constant diffusivity (thin colored lines) between 500~m$^{2}$s$^{-1}$ and 2500~m$^{2}$s$^{-1}$. Southward transports are negative, northward positive, zero transport is indicated by the thin grey line; and (b) the suppressed (solid) and unsuppressed (dashed) diffusivities, zonally averaged over the entire Southern Ocean and meridionally avereaged in the ACC lattitudes.}\label{Fig10:Eddy_Transport_Variable_vs_Constant}
\end{figure}

To further investigate the influence of the three-dimensional diffusivity on the MOC, we compare the zonally integrated eddy volume transport $\overline{v^{\prime}h^{\prime}}$ (i.e. the transport itself, not the streamfunction) computed using our diffusivity estimates, to the transport obtained assuming constant diffusivities of between 500 and 3500~m$^{2}$s$^{-1}$, meridionally averaged over the ACC envelope (Fig. \ref{Fig10:Eddy_Transport_Variable_vs_Constant}a). A similar zonal and meridional averaging is applied to the spatially variable diffusivities $K_0$ and $\widetilde{K}$ (Fig. \ref{Fig10:Eddy_Transport_Variable_vs_Constant}b). It is clear from Fig. \ref{Fig10:Eddy_Transport_Variable_vs_Constant}a that the vertical structure of the meridional transport obtained using $K_0$ resembles those obtained using constant diffusivities, with relatively strong southward transports in the ocean interior that peak at $\gamma=$26.7, 26.9 and 27.5~kg.m$^{-3}$.

In contrast, the interior southward transport determined using the suppressed eddy-diffusivity is much more modest and has a different vertical structure, reaching a peak near the critical layer, which occurs at approximately 27.8~kg.m$^{-3}$. Although the suppressed transport shows a peak transport near 26.9~kg.m$^{-3}$, similar to the unsuppressed case, the suppressed transport on this isopycnal is factor of 4 smaller in magnitude than that of the unsuppressed transport. In short, the mean flow of the ACC strongly suppress the intensity of eddy-diffusion, which dramatically reduces the southward interior geostrophic eddy-induced transport, and concentrates it in the denser water masses near the critical layer. Assuming that the simple parameterization used here is valid, it is clear that the modification of the vertical structure of the diffusivity has important implication for the Southern Ocean overturning.

%%=========% 
%%Figure 10
%%=========%
\afterpage{\clearpage}
 
\subsection{The Residual Overturning}

\begin{figure}[t]
   \centering  
  \includegraphics[width=18pc,height=20pc,angle=0]{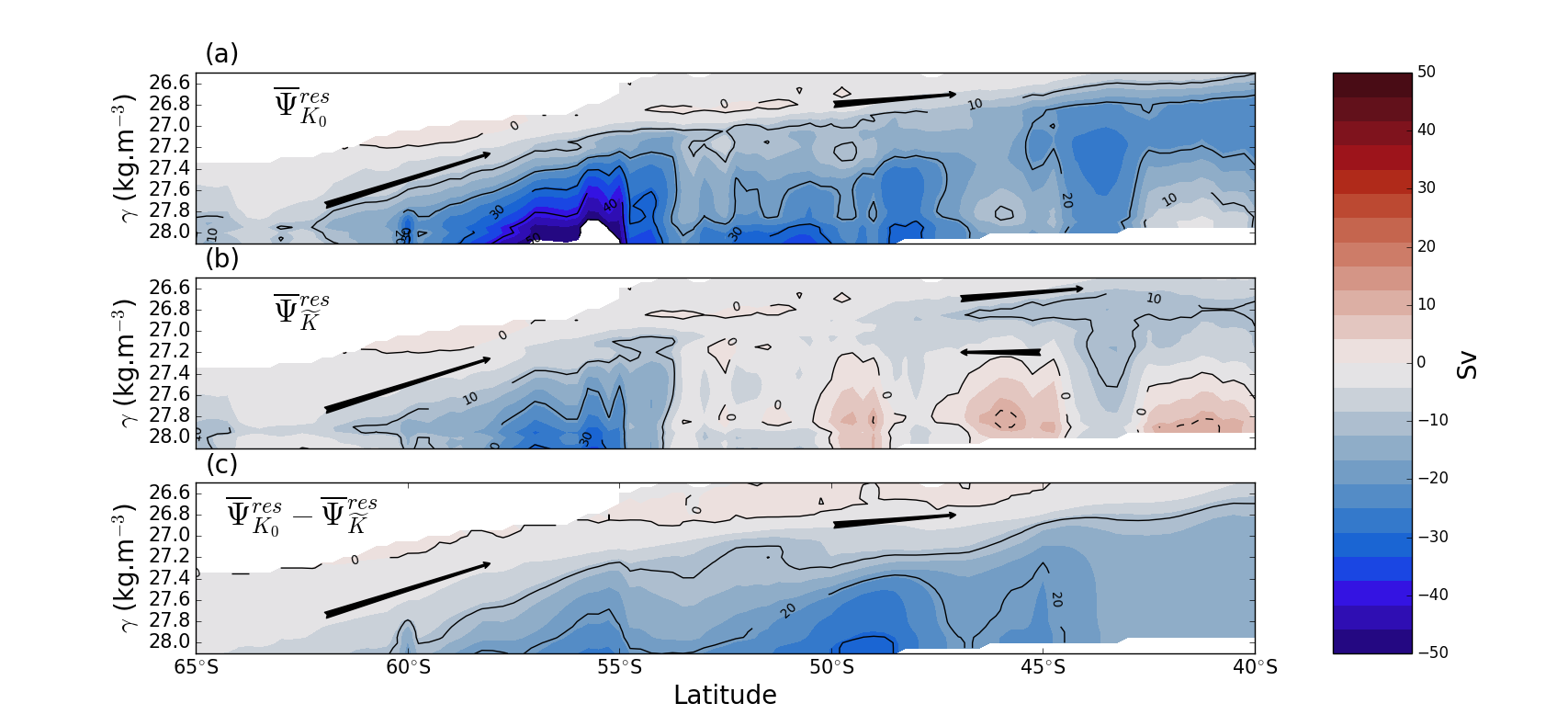}\\
  \caption{The residual overturning streamfunction $\Psi^{\textrm{res}}$ reconstructed from the observations. (a) The residual overturning streamfunction claculated using the unsuppressed diffusivity $K_0$; (b) as in \ref{Fig11:Residual_Streamfunction}a, this time calculated using the suppressed diffusivity $\tilde{K}$; and (c) the difference between the $\Psi^{\textrm{res}}_{K_{0}}$ and  $\Psi^{\textrm{res}}_{\tilde{K}}$.  Arrows indicate the sense of the overturning.}\label{Fig11:Residual_Streamfunction}
\end{figure}

With the time-mean and eddy components of the overturning in hand, we are now able to reconstruct the total residual meridional circulation (Eqn. \ref{Eqn:Meridional_Transport_GeoDecomp}) and its overturning streamfunction (Eqn. \ref{Eqn:Streamfunction_Construction}). The zonally integrated residual overturning streamfunction $\Psi^{\textrm{res}}$ is shown using the unsuppressed diffusivities in Fig. \ref{Fig11:Residual_Streamfunction}a, using the suppressed diffusivities in Fig. \ref{Fig11:Residual_Streamfunction}b, and the difference between them in Fig. \ref{Fig11:Residual_Streamfunction}c. 

Firstly, we note that our estimated residual overturning streamfunctions show numerous features in common with those computed from sophisticated numerical models \citep{DufourEtAl2012,Mazloff2013,ZikaEtAl2013}. For both the suppressed and unsupressed diffusivities, the residual overturning streamfunction is generally clockwise, with upwelling in the Drake Passage latitudes and northward flow in the lighter water masses near the surface. Although our dataset does not sufficiently sample waters denser than about 28~kg.m$^{-3}$ and thus cannot resolve the northward abyssal cell, there is some suggestion of northward flow closing the clockwise cell at about 60$^{\circ}$S, and in the blocked latitudes north of about 50$^{\circ}$S. We find a peak overturning transport of about 60~Sv when using the unsuppressed diffusivities, and a more realistic 30~Sv when using the suppressed diffusivities. Unsurprisingly, the clockwise overturning cell is much stronger with unsuppressed diffusivities, and the northward flow is found in denser (deeper) levels, as can be seen in Fig. \ref{Fig11:Residual_Streamfunction}c.

%%=========%
%%Figure 11
%%=========%

Although our reconstructions show numerous realistic features, it is clear that a perfect reconstruction eludes us. One illustration of this imperfection is that our residual streamfunction is quite noisy, although it should be noted that \cite{Mazloff2008} also produced a noisy residual streamfunction when attempting a reconstruction from each of the individual components from SOSE model output. When comparing our results with the SOSE reconstruction \citep{Mazloff2008,Mazloff2013}, we find that when using the suppressed diffusivities, the clockwise deep overturning cell is weaker in the region to the north of Drake Passage, and that the zero Sverdrup transport line that separates the deep overturning cell from the Antarctic Bottom Water (AABW) cell is shallower in our observation-based estimate, contrary to the overturning obtained using the unsuppressed diffusivities where the southward cell is stronger and deeper than the SOSE-based estimate. It is likely that the eddy volume flux of the present study estimated using the unsuppressed diffusivities is too strong, while that estimated using the suppressed diffusivities is too weak. However, we note that the strong decrease of the eddy flux that occur when using a suppressed diffusivity leads to a residual overturning in closer agreement with the results of numerical models, particularly in the Drake Passage latitudes.

\begin{figure*}[t!]
   \centering  
  \includegraphics[width=40pc,height=20pc,angle=0]{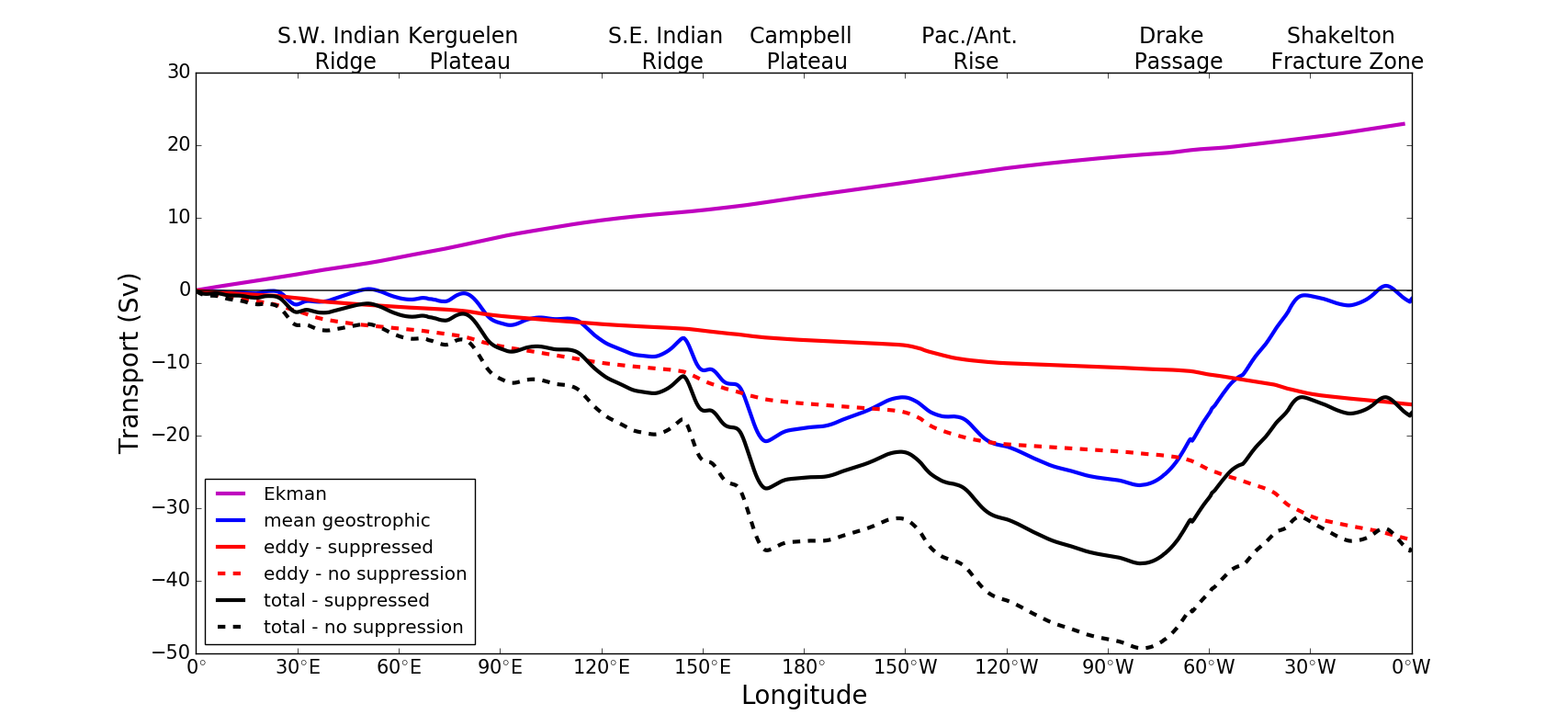}\\
  \caption{The longitudinal variation of the vertically integrated cumulative meridional transport and each of its components, averaged over the ACC latitudes. The ageostrophic Ekman transport $\overline{v}_{\textrm{EK}}\overline{h}$ (magenta); the mean geostrophic transport $\overline{h}\overline{v}_{\textrm{g}}$, the geostrophic eddy transport $\overline{v_{\textrm{g}}^{\prime}h^{\prime}}\approx K \frac{\overline{h}}{\overline{q}}\nabla\overline{q}$ (red) for the suppressed (solid) and unsuppressed (dashed) diffusivities; and the total transport $\overline{vh}=\overline{v}_{\textrm{EK}}\overline{h} + \overline{h}\overline{v}_{\textrm{g}} + \overline{v_{\textrm{g}}^{\prime}h^{\prime}}$ (black) for the suppressed (solid) and unsuppressed (dashed) diffusivities. }\label{Fig12:Cumulative_Transport}
\end{figure*}

We can explore the distribution of the meridional volume transport throughout the Southern Ocean by calculating the vertically integrated cumulative transport (determined by integrating the transport along lines of constant latitude) for each of the contributing components, averaged over the ACC envelope, shown in Fig. \ref{Fig12:Cumulative_Transport}. Here, we can gauge the influence of the Southern Ocean's bathymetry on the meridional transport, as well as see how the diffusivity suppression influences the transport around the Southern Ocean. We see that, similarly to the cross-frontal transport computed from a high-resolution numerical model by \cite{DufourEtAl2015}, the time-mean geostrophic transport (solid blue line) is concentrated in regions close to large bathymetric features, with step-changes in the transport near the Kerguelen Plateau, the Campbell Plateau, and the Pacific Antarctic Rise (indicated in Fig. \ref{Fig12:Cumulative_Transport}). As discussed previously, the mean geostrophic transport is primarily southward, balancing the northward Ekman transport (solid magenta line). However, there is a large northward mean transport that occurs as the ACC passes through Drake Passage, largely associated with the strong western boundary current that forms along the coast of South America. This northward transport balances the majority of the accumulated southward transport, resulting in effectively zero total time-mean geostrophic transport. 

%%=========%
%%Figure 12
%%=========%

In contrast, both the suppressed and unsuppressed eddy transport show a relatively uniform southward transport throughout the Southern Ocean, with a series of step changes of enhanced southward transport near certain bathymetric features, most clearly seen in the unsuppressed transport\footnote{We note that if the eddy transport is computed across the mean streamlines as opposed to meridionally, the constant southward transport disappears and the eddy transport instead manifests as a series of step changes.}. The eddy volume flux concentrated in the step-changes downstream of large bathymetric features corresponds to the locations ``storm tracks'' or ``mixing hot spots'' identified by previous studies \citep{Thompson&Sallee2012,DufourEtAl2015,ChapmanEtAl2015}. However, these concentrated regions of southward transport are generally limited in magnitude, being between 2-5~Sv for the unsuppressed diffusivities and 5-10~Sv for the suppressed diffusivities.

\afterpage{\clearpage}
\section{Influence of a Vertically Varying $K$ in a Simple Conceptual Model} \label{Section:Numerical_Model}	

In lieu of a constant diffusivity $K$, we employ in Eqn. \ref{Eqn:TEM_3} a diffusivity with a simple vertically varying structure: 
\begin{equation} \label{Eqn:Model_K_Definition}
\widetilde{K}(z) = K_0 + K^{\star}e^{-\frac{\left(z-z_c\right)^{2}}{2h_{K}}}.
\end{equation}
In this form, the diffusivity is enhanced at depth, with a peak amplitude of $K^{\star}$ at the critical level $z_c$. The vertically varying part of $K$ is superposed over a constant background diffusivity $K_0$, reminiscent of the diffusivity calculated from the observations (see Fig. \ref{Fig5:Diffusivity_With_Depth}b). The model is run over a broad parameter space, with critical layers ranging from 750m to 2000m depth, and peak $K^{\star}$ from 500 to 3500~m$^{2}$.s$^{-1}$, a similar range to those suggested by \cite{AbernatheyEtAl2010}. The background diffusivity $K_0$ is set to 250~m$^{2}$.s$^{-1}$ and the vertical scale, $h_{K}$, is set to 500~m. Additionally, we run the model with a constant, vertically invariant $K$, ranging from 500 to 3500m$^{2}$.s$^{-1}$. 

\begin{figure}[t]
   \centering  
  \includegraphics[width=20pc,height=18pc,angle=0]{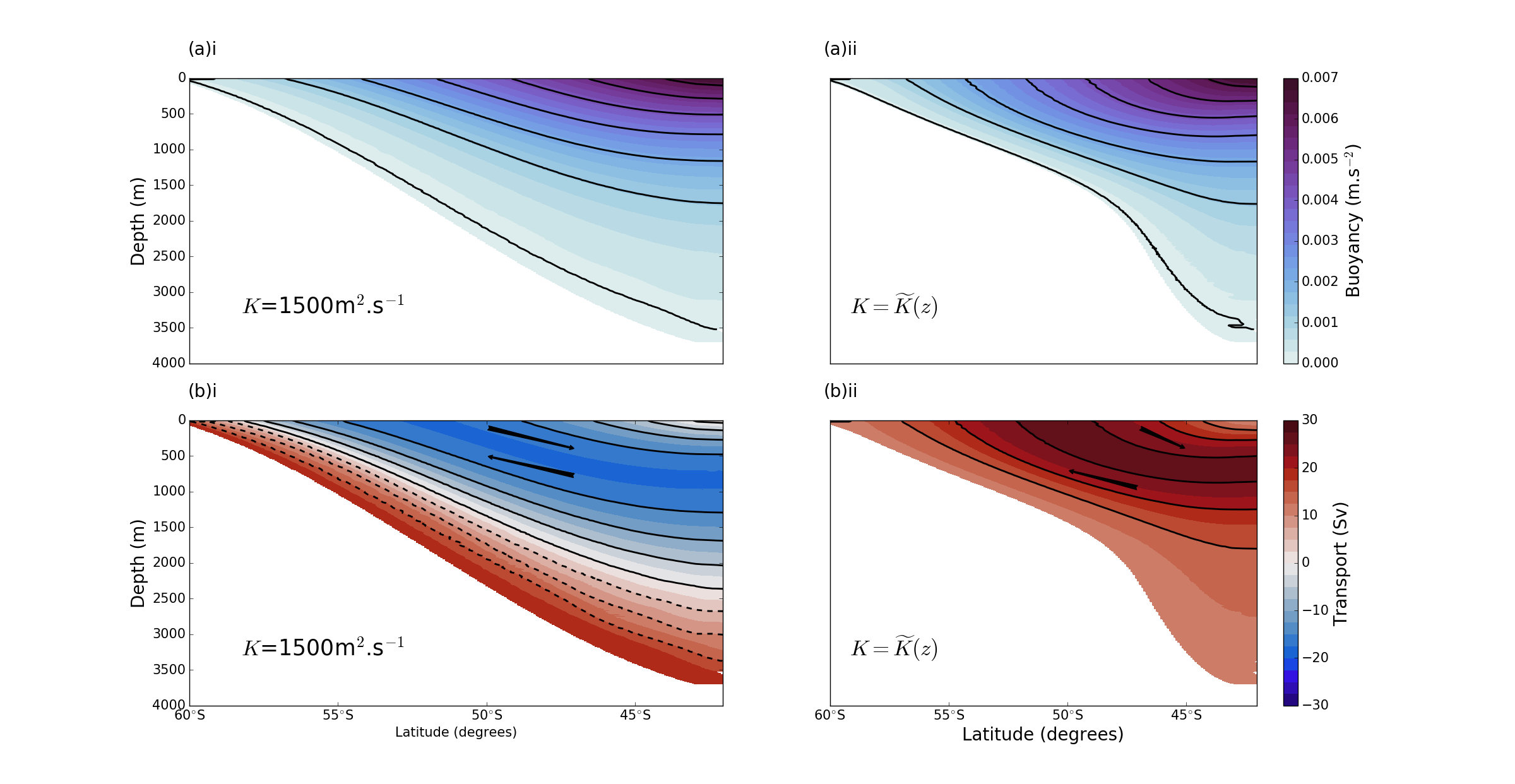}\\
  \caption{The effect of a vertically varying diffusivity on the stratification and overturning in a conceptual model. The simulated zonally averaged buoyancy field (a); and residual overturning streamfunction $\Psi^{\textrm{res}}$ (b) from the TEM model for a case with (i) a constant eddy diffusivity $K=1500$~m.s$^{-2}$; and (ii) vertically varying eddy diffusivity, given by Eqn. \ref{Eqn:Model_K_Definition} with a maximum diffusivity of $K^{\star}=1500$~m.s$^{-2}$. Solid black lines indicate contours of (a) buoyancy  (CI: 1.0$\times$10$^{-3}$~m.s$^{-2}$); and (b) overturning streamfunction (CI: 10Sv). The black arrows indicate the sense of the overturning. }\label{Fig13:TEM_output_var_vs_const}
\end{figure}

Fig. \ref{Fig13:TEM_output_var_vs_const} contrasts the results of the TEM model using a constant eddy diffusivity $K$=1500~m$^{2}$.s$^{-1}$ and using the vertically varying $K$ with peak amplitude $K^{\star}$ of 1500~m$^{2}$.s$^{-1}$. Both the buoyancy field (Fig. \ref{Fig13:TEM_output_var_vs_const}a) and the residual overturning streamfunction (Fig. \ref{Fig13:TEM_output_var_vs_const}b) show the same basic structure for the constant diffusivity (Fig. \ref{Fig13:TEM_output_var_vs_const}i) and the vertically varying diffusivity (Fig. \ref{Fig13:TEM_output_var_vs_const}ii), but with some important differences. Principally, the isopycnal inclination is greater in the case with vertically varying $\widetilde{K}$ than with constant $K$. Secondly, the maximum  $\Psi^{\textrm{res}}$ is about 10~Sv larger in the vertically varying case. Since the time-mean overturning is identical in both cases, we must conclude that the opposing eddy-overturning is weaker for vertically varying $\widetilde{K}$, despite the increased isopycnal tilt that should, by Eqn. \ref{Eqn:TEM_2}, lead to a higher eddy volume fluxes. As such, it seems that the principle result of the suppression of the eddy-diffusivity in the ocean interior is to reduce the eddy induced overturning, which in turn results in a steeper isopycnal slope.

%%=========%
%%Figure 13
%%=========%

\begin{figure*}[b]
   \centering  
  \includegraphics[width=40pc,height=20pc,angle=0]{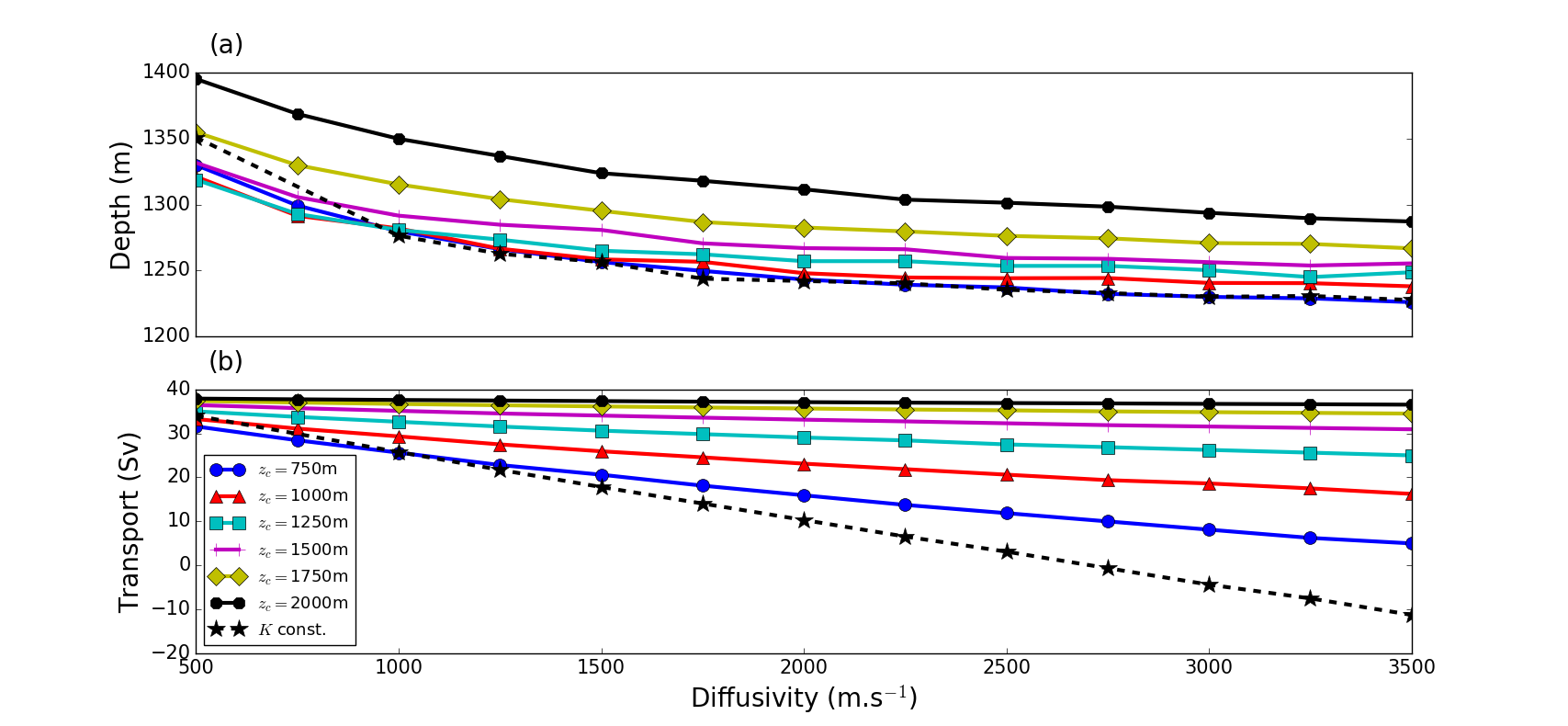}\\
  \caption{The response of the stratification (a) and the overturning streamfunction (b) to changes in the critical level depth and the eddy diffusivity $K^{\star}$. (a) The depth of the 0.2~m.s$^{-1}$ buoyancy surface at 50$^{\circ}$S; and (b) the maximum residual overturning transport. Each solid curve corresponds to simulations run with with a different critical layer depth $z_c$ in Eqn. \ref{Eqn:Model_K_Definition} (see legend in Fig. \ref{Fig14:TEM_Output_Transport_Depth_with_K}b). The dashed black curve (with `$\star$') corresponds to simulations run with a constant eddy diffusivity. }\label{Fig14:TEM_Output_Transport_Depth_with_K}
\end{figure*}

To underline this point further, Fig. \ref{Fig14:TEM_Output_Transport_Depth_with_K} shows the influence of the varying the critical layer $z_c$ and the peak eddy diffusivity $K^{\star}$ on the stratification and the residual overturning, and how the results using a vertically varying diffusivity differ from those with a constant diffusivity. Fig. \ref{Fig14:TEM_Output_Transport_Depth_with_K}a shows the depth of a representative isopycnal (in this case $\overline{b}$=0.2~m.s$^{-2}$, which is found at 1200m depth on the northern boundary) at approximately 45$^{\circ}$S. It can be seen clearly that this isopycnal shallows with increasing diffusivity and that the shallowing of isopycnals appears to approach a limit with increasing $K$. It can also been seen in Fig. \ref{Fig14:TEM_Output_Transport_Depth_with_K}a that as the critical layer depth increases (solid colored curves), the depth of the isopycnal also increases. When $K$ is constant (dashed curve) the isopycnal depth tracks closely the curve associated with the shallowest critical layer considered here (750~m), except at low values of $K$. As the critical layer deepens, the diffusivity ``felt" on this isopycnal (who's depth is constrained to be 1200~m on the northern boundary), at this latitude, increases, resulting in a flattening of the isopycnal. The exact response of the isopycnal depth depends on the choice of isopycnal, and where it lies in relation to the critical layer. As such, changes in the $z_c$ can modify vertical density gradient in the interior and, hence, the thickness of isopycnal layers.       
% * <jean-baptiste.sallee@locean-ipsl.upmc.fr> 2016-10-26T13:41:07.318Z:
%
% > It can also been seen in Fig. \ref{Fig14:TEM_Output_Transport_Depth_with_K}a that as the critical layer depth increases (solid colored curves), the depth of the isopycnal also increases.
%
% Why is this so? Presumably because the diffusivity felt at b=0.2 ms-2 reduces as the critical layer deepens. This must depends on the choice of the b surface, presumably  b=0.2 ms-2 is shallower than 750 m at 45S. If you chose a b-surface that was at 1500 m at 45S, you would expect that from z-c = 750->1500 m, diffusivity at the b-surface increase, therefore b shallows; then from z_c = 1500 ->2000m,  diffusivity at the b-surface decrease, therefore b shallows. Is that correct ?
%
% ^.

The residual overturning streamfunction is also sensitive to changes in $z_c$ and $K$. Fig. \ref{Fig14:TEM_Output_Transport_Depth_with_K}b shows the maximum value of the residual streamfunction at the northern boundary for each model run. Here, we note that in general, the overturning streamfunction decreases approximately linearly with increasing $K$ for both the vertically varying $K$ cases and the constant $K$ case. Indeed, weaker eddy diffusivity reduces the efficiency of the eddy transport to counterbalance the mean transport, which results in higher residual overturning. However, unlike the isopycnal depth shown in Fig. \ref{Fig14:TEM_Output_Transport_Depth_with_K}a, the sensitivity of the overturning transport to changes in $K$ varies depending on the critical layer depth  $z_c$. For example, with a deep critical layer, of $z_c$=2000~m (black solid line in Fig. \ref{Fig14:TEM_Output_Transport_Depth_with_K}b), the slope of the line is approximately -5$\times$10$^{-4}$~Sv/(m$^{2}$.s$^{-1}$), indicating almost no sensitivity to changes in $K$, while with a shallow critical layer of $z_c$=750~m, the streamfunction is highly sensitive to changes in $K$: the slope is approximately -1$\times$10$^{-1}$~Sv/(m$^{2}$.s$^{-1}$), almost 3 orders of magnitude higher than than when $z_c$=2000~m. As with the isopycnal depth, the constant $K$ case shows the strongest similarity with the vertically varying $K$ at the shallowest critical level, although the constant $K$ shows a steeper curve that, at high diffusivities, results in the eddy overturning dominating the time-mean overturning and a reversal of the overturning sense. 

%%=========%
%%Figure 14
%%=========%

As in the observational part of our study, the principle effect of the introducing a vertical structure to the eddy diffusivity is to suppress the 
southward interior transports by eddy-fluxes. However, the details of this effect of this suppression on the resulting overturning depend critically on the depth of the critical layer, where the diffusivity is still large enough to enable substantial eddy fluxes. For example, when the critical layer is very deep ($z_c=$2000~m), the enhanced diffusivity is deeper than the depths with substantial isopycnal slopes and hence the eddy induced transport is weak. When the critical layer is shallower, say at 750m or 1000m, as found in our observations, the enhanced diffusivity is found at levels with large isopycnal slopes or IPV gradients and a substantial eddy overturning can be supported, and hence the resultant residual overturning is sensitive to changes in the value of $K$. As the critical layer depth varies throughout the southern ocean \citep{Smith&Marshall2009,AbernatheyEtAl2010,ColeEtAl2015}, this result could have important implications for the local eddy-flux and its parameterization in climate models.  
\afterpage{\clearpage}

\section{Discussion and Conclusions} \label{Sec:Conclusion}

In this study, we have investigated how the three-dimensional structure of the eddy diffusivity and its suppression by the time-mean flow, can influence the MOC in the Southern Ocean. Combining hydrographic observations obtained with  satellite altimetry, we have estimated the isopycnal eddy diffusivity, $K$, using the framework of \cite{Ferrari&Nikurashin2010}, in three-dimensions and including the effect of suppression by the time-mean flow. We obtain a $K$ field that is highly spatially variable. Large values of diffusivity are found in regions downstream of large topographic features, and $K$ is suppressed in regions of strong time-mean flow. When suppression is taken into account, the diffusivity $K$ reaches a peak at the critical layer, which we find to be at about 1000~m.  Using the estimate of the eddy diffusivity,  we are able to estimate the eddy volume flux on an isopycnal as a downgradient diffusion of isopycnal potential vorticity. Additionally, using the approximate isopycnal streamfunction of \cite{McDougall&Klocker2010},  we are able to estimate the time-mean geostrophic meridional circulation. Together with the ageotrophic Ekman transport, we are able to reconstruct the full upper ocean meridional circulation.  

We have focused on the effect of the suppression of $K$ by the mean-flow on the resulting overturning. By comparing our reconstructions of the overturning circulation with, and without, the effects of the time-mean flow suppression, we are able to show that the primary effect of the suppressed diffusivity is to dramatically reduce the interior eddy flux, particularly in the intermediate and upper-circumpolar deep waters. Reconstructing the eddy overturning using either the unsuppressed diffusivity, or a constant diffusivity, strongly overestimates these interior volume fluxes (at least when compared to the output from the eddy-permitting SOSE model, described in \cite{Mazloff2013}). We find that the parameterized eddy fluxes, as well as the time-mean geostrophic flows, are zonally asymmetric, being concentrated near, or downstream, of bathymetric features, in regions corresponding to the mixing ``hot spots'' or ``storm tracks'' identified in previous studies.  

One inherent limitation of our observation-based approach is that diffusivity and stratification are intrinsically related, while here we apply differing diffusivity on a fixed stratification. To go beyond this limitation and further explore how the overturning responds to the depth varying structure of the eddy diffusivity, we use a simple conceptual model of the Southern Ocean, based on that of \cite{Marshall&Radko2003} and \cite{Marshall&Radko2006}. We find that, as in the observational part of this study, the addition of a vertical structure to the eddy diffusivity acts to suppress the interior southward eddy transport when compared to model runs performed using a vertically constant $K$. The resulting stratification and overturning circulation is also sensitive to the depth of the critical layer. As the critical layer becomes shallower, the overturning transport becomes more sensitive to changes in the magnitude of the peak diffusivity, as the critical layer and its associated region of high eddy diffusivities, is more likely to coincide with regions with large isopycnal slopes or potential vorticity gradients.

The principle result of this study is that the mean flow of the Antarctic Circumpolar Current is critical in shaping the interior Southern Ocean overturning circulation, not only driving a significant time-mean geostrophic overturning, a point emphasized by \cite{Mazloff2008} and \cite{Mazloff2013}, but also modulating the efficiency the resulting eddy overturning circulation. We find that the details of the overturning and interior stratification are sensitive to both the magnitude of $K$, and also the depth of the critical layer, which both depend on a subtle balance between eddy characteristics and mean-flow. The corollary of this result is that in order to reconstruct an overturning circulation using a downgradient parameterisation, correctly representing the interior suppression of eddy diffusivity by the mean-flow is crucial. In contrast, the zonal asymmetry, although important for the localisation of the volume transport, is of second order importance when considering the zonally averaged circulation, as revealed by the fact that the structure of the zonally averaged meridional volume flux computed using the (spatially varying) unsuppressed diffusivity is similar to that obtained using constant values of the $K$ (see Fig. \ref{Fig10:Eddy_Transport_Variable_vs_Constant}). The fact that the vertical structure of the diffusivity plays such an important role in parameterized eddy flux may have important implications for coarse-resolution ocean models used for long-period climate studies, as these models still rely on downgradient turbulence closures such as Gent-McWilliams. Further research will further explore the role of the vertical diffusivity structure in the response to climate change, as well refining our estimate of the overturning circulation through the use of new data, parameterizations and analysis techniques.

%Climate change is expected to increase not only the wind stress over the Southern Ocean, but also eddy activity \citep{Meredith&Hogg2006,MorrowEtAl2010} and there is some limited evidence that the strength of the eddy field in the Southern Ocean may have increased over the last 20 years in certain regions \citep{HoggEtAl2014}. The response of the overturning to any changes in surface wind stress will depend on not only on the magnitude of the eddy response but also the response of the background mean-flow and the vertical structure of the eddy field. For example, there is some evidence that climate change will lead to an increase in the stratification of the Southern Ocean \citep{MeijersEtAl2012}, which could, in turn, result in a shallower penetration of the ACC into the interior and corresponding shallowing of the critical layer, which could induce increased interior eddy fluxes. Further research will further explore the role of the vertical diffusivity structure in the response to climate change, as well refining our estimate of the overturning circulation through the use of new data, parameterizations and analysis techniques.   

%%%%%%%%%%%%%%%%%%%%%%%%%%%%%%%%%%%%%%%%%%%%%%%%%%%%%%%%%%%%%%%%%%%%%
% ACKNOWLEDGMENTS
%%%%%%%%%%%%%%%%%%%%%%%%%%%%%%%%%%%%%%%%%%%%%%%%%%%%%%%%%%%%%%%%%%%%%
\section*{Acknowledgments} 
The authors thank Dhruv Balwada, Jessica Masich, Andreas Klocker and Christopher Roach for useful discussions and to Christopher Roach for helpfully providing the diffusivity data from \cite{RoachEtAl2016} for comparison with our calculations. CC was supported by an NSF Division of Ocean Sciences postdoctoral fellowship Grant No. 1521508. J.B.S. received support from Agence Nationale de la Recherche (ANR), ANR-12-PDOC-0001.

%%%%%%%%%%%%%%%%%%%%%%%%%%%%%%%%%%%%%%%%%%%%%%%%%%%%%%%%%%%%%%%%%%%%%
% APPENDIXES
%%%%%%%%%%%%%%%%%%%%%%%%%%%%%%%%%%%%%%%%%%%%%%%%%%%%%%%%%%%%%%%%%%%%%

%% If only one appendix, use
%\appendix%

%% If more than one appendix, use \appendix[<letter>], e.g.,
% \appendix[A] 
\section*{Appendix A: Data Availability}
All interpolated fields used in this study, including the estimates of the suppressed and unsuppressed eddy diffusivity; the neutral density, the approximate isopycnal geostrophic streamfunction and its variance; and the isopycnal potential vorticity, are available for download in NetCDF format from the following URL:

The output is provided annually. Additionally, the first two seasonal harmonics are estimated and are included in the output.
\section*{Appendix B: Numerical Method for the Conceptual Model}
The primary equation that needs to be solved for the implementation of the conceptual TEM model (Eqn. \ref{Eqn:TEM_3}) has the form:
\begin{equation} \label{Eqn:A1}
A(y,b,\Psi^{\textrm{res}})\frac{\partial \overline{b}}{\partial y} + B(y,b,\Psi^{\textrm{res}})\frac{\partial \overline{b}}{\partial z} = 0
\end{equation}
where $A$ and $B$ are coefficients that are functions of the surface wind stress and the residual overturning streamfunction, together with Dirichlet boundary conditions for $\overline{b}$ at $z=$ and $y=L_y$. Using the method of characteristics, this linear partial differential equation can be written as the set of coupled ordinary differential equations:
\begin{eqnarray} \label{Eqn:A2}
\frac{d y}{d s} & = & A(y,\overline{b},\Psi^{\textrm{res}}) \\ \label{Eqn:A3}
\frac{d z}{d s} & = & B(y,\overline{b},\Psi^{\textrm{res}}) \\ \label{Eqn:A4}
\frac{d \overline{b}}{d s} & = & 0 \\
\frac{d \Psi^{\textrm{res}}}{ds} & = & 0
\end{eqnarray}
where $s$ is the distance along the characteristic curve which, in this case, is simply the isopycnal $\overline{b}$, together with the boundary conditions:
\begin{eqnarray} \label{Eqn:A5}
\overline{b}(y,z=0) & = & g(y) \\ \label{Eqn:A6}
\overline{b}(y=L_y,z) & = & f(y)
\end{eqnarray}
Eqns. \ref{Eqn:A2}--\ref{Eqn:A4} are solved using a 4th order Runge-Kutta method. Boundary conditions are imposed using the shooting method: using large initial guesses of $\Psi^{\textrm{res}}=\pm$100Sv, and starting at $z=0$, Eqns. \ref{Eqn:A2}--\ref{Eqn:A4} are integrated until $y=L_y$. We then compare the depth of the isopycnal to the depth of that isopycnal expected from the boundary conditions Eqn. \ref{Eqn:A6} and apply the bisection method to systematically adjust the guess of $\Psi^{\textrm{res}}$ until convergence to a predefined error tolerance (here, 5m).  

Computer code to implement this model, written in the open-source, Python programming language, is available under an open source MIT license from CC's Github account:
https://github.com/ChrisC28

%----------------------------------------------------------------------------------------
%	REFERENCE LIST
%----------------------------------------------------------------------------------------

%----------------------------------------------------------------------------------------

\end{document}